\documentclass[twocolumn,prx,preprintnumbers,superscriptaddress,amsmath,amssymb,english]{revtex4}
\usepackage{amsmath,amssymb,amsfonts,mathrsfs}
\usepackage{amsthm}
\usepackage{CJK}
\usepackage{bm}
\usepackage{lipsum}
\usepackage{babel}
\usepackage{graphicx}
\allowdisplaybreaks
\usepackage{makecell}
\usepackage{booktabs}
\usepackage{multirow}
\usepackage{gensymb}
\usepackage{rotating}
\usepackage{float}
\usepackage{braket}
\usepackage[breaklinks]{hyperref}
\usepackage{color}
\hypersetup{colorlinks=true, linkcolor=blue, citecolor=blue, filecolor=blue, urlcolor=blue}
\begin{document}

\title{Controllable Weyl nodes and Fermi arcs in a light-irradiated carbon allotrope}

\author{Ruoning Ji}
\affiliation{Institute for Structure and Function $\&$ Department of Physics $\&$ Chongqing Key Laboratory for Strongly Coupled Physics, Chongqing University, Chongqing 400044, P. R. China}

\author{Xianyong Ding}
\affiliation{Institute for Structure and Function $\&$ Department of Physics $\&$ Chongqing Key Laboratory for Strongly Coupled Physics, Chongqing University, Chongqing 400044, P. R. China}

\author{Fangyang Zhan}
\affiliation{Institute for Structure and Function $\&$ Department of Physics $\&$ Chongqing Key Laboratory for Strongly Coupled Physics, Chongqing University, Chongqing 400044, P. R. China}

\author{Xiaoliang Xiao}
\affiliation{Institute for Structure and Function $\&$ Department of Physics $\&$ Chongqing Key Laboratory for Strongly Coupled Physics, Chongqing University, Chongqing 400044, P. R. China}

%\author{Junjie Zeng}
%\affiliation{Institute for Structure and Function $\&$ Department of Physics $\&$ Chongqing Key Laboratory for Strongly Coupled Physics, Chongqing University, Chongqing 400044, P. R. China}

\author{Jing Fan}
\affiliation{Center for Computational Science and Engineering, Southern University of Science and Technology, Shenzhen 518055, P. R. China}

\author{Zhen Ning}
\email[]{zhenning@cqu.edu.cn}
\affiliation{Institute for Structure and Function $\&$ Department of Physics $\&$ Chongqing Key Laboratory for Strongly Coupled Physics, Chongqing University, Chongqing 400044, P. R. China}

\author{Rui Wang}
\email[]{rcwang@cqu.edu.cn}
\affiliation{Institute for Structure and Function $\&$ Department of Physics $\&$ Chongqing Key Laboratory for Strongly Coupled Physics, Chongqing University, Chongqing 400044, P. R. China}
\affiliation{Center of Quantum materials and devices, Chongqing University, Chongqing 400044, P. R. China}

\begin{abstract}
The precise control of Weyl physics in realistic materials offers a promising avenue to construct accessible topological quantum systems, and thus draw widespread attention in condensed-matter physics. Here, based on first-principles calculations, maximally localized Wannier functions based tight-binding model, and Floquet theorem, we study the light-manipulated evolution of Weyl physics in a carbon allotrope C$_ 6$ crystallizing a face-centered orthogonal structure (fco-C$_6$), an ideal Weyl semimetal with two pairs of Weyl nodes, under the irradiation of a linearly polarized light (LPL). We show that the positions of Weyl nodes and Fermi arcs can be accurately controlled by changing light intensity. Moreover, we employ a low-energy effective $k\cdot p$ model to understand light-controllable Weyl physics. The results indicate that the symmetry of light-irradiated fco-C$_6$ can be selectively preserved, which guarantees that the light-manipulated Weyl nodes can only move in the high-symmetry plane in momentum space. Our work not only demonstrates the efficacy of employing periodic driving light fields as an efficient approach to manipulate Weyl physics, but also paves a reliable pathway for designing accessible topological states under light irradiation.
\end{abstract}


%Use showkeys class option if keyword

\maketitle
\section{Introduction}
The extension of topological insulators (TIs) to topological semimetals (TSMs) has provided excellent avenues to study topological phases of matter \cite{Kane-RevModPhys.82.3045, ZSC-RevModPhys.83.1057,RevModPhys.90.015001}. In comparison with gapped TIs, TSMs possess the nontrivial gapless electronic excitations near the Fermi level. According to the degeneracy features between valence and conduction bands in the momentum space, several TSMs have been classified. The typical representatives are Dirac semimetals \cite{WangPhysRevB.85.195320, ScienceLiu864}, Weyl semimetals(WSMs) \cite{Wan2011,Xu2011,PhysRevX.5.011029,Xu2015}, and nodal line semimetals \cite{PhysRevB.84.235126,PhysRevLett.115.036806,PhysRevLett.115.036807}. The fermionic quasiparticles in these TSMs have analogues of relativistic fermions in the quantum field theory. Besides, since low-energy quasiparticles in crystalline materials are constrained by the discrete crystal symmetry rather than the Poincar\'{e} symmetry, more unconventional quasiparticles without high-energy physics counterparts, such as type-II Weyl fermions \cite{NatureSoluyanov2015}, triple fermions \cite{ZhuPhysRevX.6.031003,lvNature2017}, hourglass fermions \cite{Wang2016, PhysRevLett.123.126403, Hourglass}, as well as beyond \cite{Bradlynaaf5037}, have been present in TSMs.

Among various TSMs, WSMs with fermionic excitations depicted by two-component Weyl equations around the band crossing nodes (i.e., Weyl nodes) are of widespread interest. Due to the twofold degeneracy of Weyl nodes,  breaking either time-reversal or inversion symmetry must occur in WSMs. The topology of Weyl nodes is featured by a quantized chiral charge, acting as "magnetic monopoles" of Berry curvature in the momentum space \cite{Wan2011,Xu2011,Sciencemonople2003}. %Consequently, a distinct hallmark of WSMs is the occurrence of nontrivial Fermi arcs that connect the projections of Weyl nodes with opposite chiral charge \cite{Wan2011}.
As WSMs possess such attractive features, the recent exploration of WSM phases in realistic materials have been intensely investigated in both theories and experiments \cite{PhysRevX.5.011029, Xu2015, Wang2017, Wang20172, WangPRL2019, Morali1286, Belopolski1278, Liu1282, PhysRevLett.127.277204}. Another important but generally ignored issues of WSMs is the manipulation of Weyl nodes. In realistic WSMs, the positions of Weyl nodes are associated with the crystalline symmetry and chemical compositions, and thus Weyl nodes are fixed at specific positions in the momentum space and slightly influenced by perturbations. As the separation of a pair of Weyl nodes intimately relates to the fundamental properties of WSMs, one needs to explore reliable approaches to control the positions of Weyl nodes. Recently, periodic driving via light irradiation offers a fascinating avenue to realize Floquet-Bloch states in condensed systems \cite{Shirley,Goldman,Eckardt2015,Bukov2015,Oka2019, Light2022}. Through changing the propagation or polarization direction of incident light, we can conveniently realize the symmetry modification and control band structures, and thereby engineer desired Floquet topological phases with highly tunability \cite{exp1,exp2,exp3}. Over the last decade, various optically driven topological states with exotic features have been intensely studied \cite{FT1,FT2,Rev1,Rev2,Rev3}. Strikingly, WSM phases have been predicted in several light-irradiated nodal line and Dirac semimetallic phases \cite{FWSM1,FWSM2,FWSM3,FWSM4,FWSM5,FWSM6,FWSM7,FWSM8,FWSM9}. The Weyl nodes in these light-induced WSMs exhibit controllable chirality or positions. While the advancements have been very encouraging, the most interesting features established so far are mainly based on low-energy effective models. The fascinating light-manipulated Weyl physics and their emergent Fermi arcs in realistic materials have been largely unexplored.

\begin{figure}
    \centering
    \includegraphics[width=\linewidth]{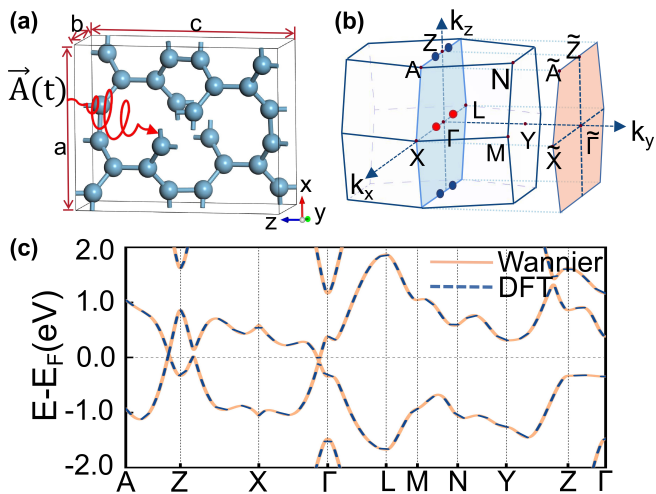}
   \caption{(a) Schematic diagram of carbon allotrope fco-C$_6$ irradiated by the periodic driving light field $\mathbf{A}(t)$. The lattice parameters $a$, $b$, and $c$ of fco structures are also denoted. (b) The bulk BZ of fco structures and the corresponding (010)-surface BZ. The Weyl nodes with chirality $C=+1$ and $C=-1$ are denoted by red and blue solid dots. (c)  The comparison of electronic band structures of fco-C$_6$ along high-symmetry path obtained from first-principles calculations (red solid-lines) and MLWF-based TB Hamiltonian (black dashed-lines).
    \label{FIG1}}
\end{figure}

In this work, we study the light-manipulated Weyl nodes in a carbon allotrope (i.e., fco-C$_6$) under periodic light irradiation. The carbon allotrope fco-C$_6$ was previously demonstrated to host two pairs of symmetry-protected Weyl nodes \cite{PhysRevB.101.235119}. As shown in Fig. \ref{FIG1}(a), the carbon allotrope fco-C$_6$ crystallizes in a noncentrosymmetric face-centered orthogonal (fco) structure with a space group $F222$ (No. 22), denoted by three lattice parameters $a$, $b$, and $c$. There are six carbon atoms in one unit cell. Its bulk Brillouin-zone (BZ) and corresponding (010)-surface BZ are shown in Fig. \ref{FIG1}(b). The Weyl nodes in fco-C$_6$ lie along the high-symmetry path $\Gamma-X$ ($Z-A$) at the high-symmetry plane $k_x-k_z$ with $k_y = 0$ \cite{PhysRevB.101.235119}. Once the periodic light irradiation $\mathbf{A}(t)$ is applied, the interplay of the vector potential of light and electron momentum must destroy the specific symmetry. As a result, the positions of Weyl nodes may be away from their initial locations and can be exactly controlled by the applied light irradiation. Since the extremely weak spin-orbital coupling (SOC) effect of carbon element leads to the negligible interaction between the SOC and light irradiation \cite{PhysRevB.102.201105},  fco-C$_6$ can be considered as a perfect platform to study light-manipulated Weyl physics.

\section{Computational Method}
We carried out first-principles calculations based on density functional theory (DFT) \cite{PhysRev.136.B864,PhysRev.140.A1133}. The electron-ion interaction was treated by the projector augmented-wave method as encoded in the Vienna $ab$ $initio$ Simulation Package (VASP) \cite{PhysRevB.54.11169}. The plane-wave cutoff energy was set to be 800 eV. The exchange-correlation functional was described by the generalized gradient approximation within the Perdew-Burke-Ernzerhof formalism  \cite{PhysRevLett.77.3865}. The first Brillouin-zone was sampled by $12 \times 12 \times 12$ Monkhorst-Pack mesh grid.  The geometric structure of fco-C$_6$ was fully relaxed until total energy and forces on each atom were converged to $10^{-8}$ eV and 0.001 eV/\text{\AA}, respectively. To illustrate the light-manipulated Weyl nodes of carbon allotrope fco-C$_6$, we employed the maximally localized Wannier functions (MLWF) to construct MLWF-based tight-binding (TB) Hamiltonians from first-principles calculations based on DFT \cite{Marzari2012, MOSTOFI2008685}. The $s$, $p_x$, $p_y$, and $p_z$ orbitals of C atom were chosen to initialize the MLWFs, and matrix elements of MLWF-based TB Hamiltonian can be expressed as
\begin{equation}
\begin{split}
H_{mn}^{W}(\mathbf{k}) &= \big< \psi_{\mathbf{k},m}({\mathbf{r}}) | \hat{H}({\mathbf{r}}) | \psi_{\mathbf{k},n}({\mathbf{r}}) \big> \\
&= \sum_{\mathbf{R}} e^{i \mathbf{k} \cdot [\mathbf{R}-(\mathbf{\tau}_m-\mathbf{\tau}_n)]} t_{mn} (\mathbf{R} - \mathbf{0}) \\
\end{split}
\end{equation}
with
\begin{equation}
t_{mn} (\mathbf{R} - \mathbf{0}) = \big<\mathbf{0} + \mathbf{\tau}_{n} | \hat{H}({\mathbf{r}}) | \mathbf{R} + \mathbf{\tau}_{m} \big>.
\end{equation}
Here, $| \psi_{\mathbf{k}} \big>$ is Bloch wave functions with wavevector $\mathbf{k}$, $\mathbf{R}$ is the Bravais lattice vector, $\mathbf{\tau}$ denotes the internal freedom, and $t_{mn} (\mathbf{R} - \mathbf{0})$ is the hopping amplitude from Wannier orbital $n$ at home site $\mathbf{0}$ to Wannier orbital $m$ at site $\mathbf{R}$. The surface local density of states and the Fermi arcs were obtained by the iterative Green's function method using WANNIERTOOLS package \cite{WU2018405, Sancho1984}.
%Using the WANNIERTOOLS package based on the iterative Greens method \cite{WU2018405, Sancho1984}, the topological edge states such as the local density of states (LDOS), were calculated \cite{WU2018405, Sancho1984}. The DFT electronic band structures in the region of $E_{F} \pm 2$ eV as illustrated in Fig. 2 can be accurately reproduced by the Wannier-function-based tight-binding Hamiltonian with s and p orbitals of carbon.

When a monochromatic light field is applied to fco-C$_6$ [see Fig. \ref{FIG1}(a)], we can use the Peierls substitution to obtain a time-dependent Hamiltonian as
\begin{equation}
H_{mn}^{W}(\mathbf{k},t)= \sum_{\mathbf{R}} e^{i [\mathbf{k}+\frac{e}{\hbar}\mathbf{A}(t)] \cdot [\mathbf{R}-(\mathbf{\tau}_m-\mathbf{\tau}_n)]} t_{mn} (\mathbf{R} - \mathbf{0}),
\end{equation}
where $\mathbf{A}(t)$ is a time-periodic and space-homogeneous vector potential as $\mathbf{A}(t)=\mathbf{A}(t+T)$ with period $T$, and then the polarized electric field is $\mathbf{E}(t)=-\partial_{t}\mathbf{A}(t)$. In this paper, we consider the off-resonant region with frequency $\omega=2\pi/T$ that is significantly larger than the typical energy scale of system, and thereby we can have an effectively static Hamiltonian as \cite{Goldman,FWSM3}
\begin{equation}\label{Hfirstf}
H_{mn}^{\mathrm{eff}}(\mathbf{k})=H_{mn}^{0}(\mathbf{k})+\sum_{q\geq 1}\frac{[H_{mn}^{-q},H_{mn}^{q}]}{q \omega}+O(\frac{1}{\omega ^2}),
\end{equation}
with
\begin{equation}
\begin{split}
H_{mn}^{q}({\mathbf{k}}, \omega)= \frac{1}{T} \int_{0}^{T} {e}^{-\mathrm{i} q \omega t} H_{mn}^{W}(\mathbf{k},t) \mathrm{d} t.
\end{split}
\end{equation}
Here, $H_{mn}^{q}({\mathbf{k}}, \omega)$ is the $q$th-order time-independent Floquet Hamiltonian, which is generally in infinite dimensional Hilbert space. However, the matrix elements of $H_{mn}^{q}({\mathbf{k}}, \omega)$ decay rapidly with $|q|$ in the high-frequency approximation. Here, we use the effectively static Hamiltonian $H_{mn}^{\mathrm{eff}}(\mathbf{k})$ that is obtained from the second order truncation. This can well describe the photo-dressed band structures.

\begin{figure}
    \centering
    \includegraphics[width=\linewidth]{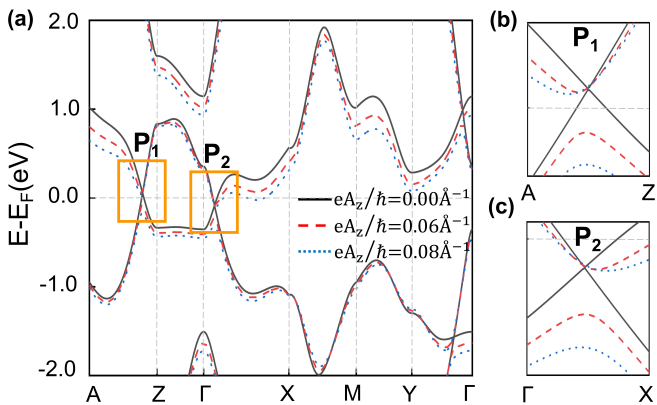}
   \caption{Evolution of electronic band structures of fco-C$_6$ under the irradiation of LPL with a light intensity $eA_z/\hbar =$ 0.00, 0.06, and 0.08 \AA$^{-1}$, respectively. The enlarged views illustrated in panels (b) and (c) indicate that initial band crossings in the $\Gamma-X$ and $Z-A$ directions are both gapped, and the band gaps are enlarged with the increase of light intensity.
    \label{FIG2}}
\end{figure}

\section{Results and Discussion}

 As shown in Fig. \ref{FIG1}(c), we can see that electronic band structures of fco-C$_6$ obtained from MLWF-based TB Hamiltonian can accurately reproduce ones computed from DFT calculations. Therefore, we can use the MLWF-based TB Hamiltonian to further investigate the evolution of band structures of fco-C$_6$ irradiated by a time-periodic and space-homogeneous monochromatic light field. Without light irradiation, the bands along the high-symmetry pathes show crossing nodes (i.e., Weyl nodes) in the $\Gamma-X$ and $Z-A$ directions. These Weyl nodes are guaranteed by the rotational symmetry $C_{2x}$ and antiunitary mirror symmetry $\tilde{M}_y = \mathcal{T}C_{2y}$ (where $T$ and $C_{2x(y)}$ respectively denotes the time-reversal symmetry and two-fold rotational symmetry with rotational axis along $x(y)$ direction) and thus located along the $k_x$ axis in the $k_x$-$k_z$ plane with $k_y=0$. The Weyl nodes with chirality $C=+1$ are positioned at ($\pm k_{x_{0}}$, $0$, $0$), and those with chirality $C=-1$ are positioned at ($\pm k_{x_{0}}$, $0$, $2\pi/c$) [see Fig. \ref{FIG1}(b)], consistent with the previous study \cite{PhysRevB.101.235119}. To explicitly reveal the light-controlled Weyl nodes in light-irradiated fco-C$_6$, we employ a linearly polarized light (LPL) whose polarization parallels to $x$-$z$ plane and incident direction is along the $y$ axis.  Here we mainly focus on the case of the LPL with polarization along the $z$-axis, and the time-periodic vector potential is $\mathbf{A}(\tau)=(0, 0, A_z\sin(\omega \tau))$, where $A_z$ is its amplitude.

\begin{figure}
    \centering
    \includegraphics[width=\linewidth]{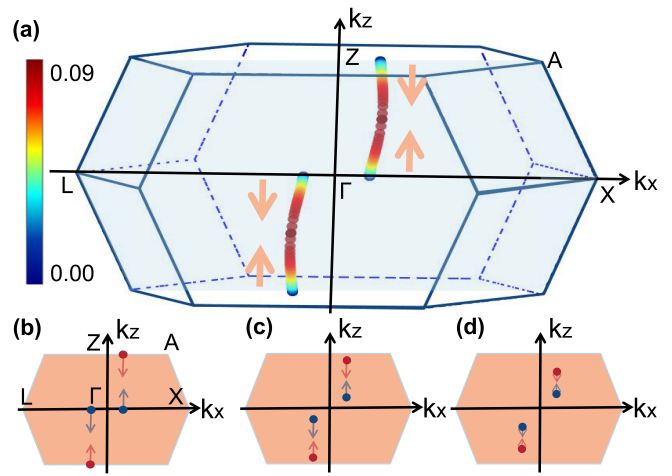}
   \caption{(a) The position of Weyl nodes as a function of light intensity highlighted in bulk BZ. The color-bar denotes the values of light intensity $eA_z/\hbar$. The distributions with a light intensity with $eA_z/\hbar =$ 0.00, 0.06, and 0.08 \AA$^{-1}$ as illustrated in panels (b), (c), and (d), respectively.
    \label{FIG3}}
\end{figure}

As depicted in Fig. \ref{FIG2}(a), we show band structures of fco-C$_6$ with different light intensities (i.e., $eA_z/\hbar =$ 0.00, 0.06, and 0.08 \AA$^{-1}$) along the high-symmetry pathes. One can find that the light-modified band structures exhibit complicated behaviors and depend on the wave-vector $\mathbf{k}$. Due to the symmetry breaking induced by the light field, the initial band crossings  in the $\Gamma-X$ and $Z-A$ directions are both gapped, and the band gaps are enlarged with the increase of light intensity [see the locally enlarged views in Figs. \ref{FIG2}(b) and \ref{FIG2}(c)]. However, Fig. \ref{FIG2}(a) indicates that the band inversion at the $\Gamma$ and $Z$ points can still be preserved, and thus the nontrivial band topology can give rise to the presence of Weyl nodes. Through searching local minimum of energy differences between valence and conduction bands, we find that, in the specific range of light intensity, there are two pairs of Weyl nodes in the bulk BZ. To illustrate the evolution of Weyl nodes, we trace the position of Weyl nodes as a function of light intensity. As shown in Fig. \ref{FIG3}, the light-manipulated Weyl nodes only move in the high-symmetry $k_x$-$k_z$ with $k_y=0$ plane. This is because our employed LPL with polarization along the $z$-axis and propagating along $y$-axis may preserve the antiunitary mirror symmetry $\tilde{M}_y$ but break the rotational symmetry $C_{2x}$. Moreover, with the increase of light intensity, we can see that the each pair of Weyl nodes move close to each other. When the light intensity exceeds a threshold value, the Weyl nodes with positive and negative chiralities annihilate and thereby the nontrivial band topology vanishes. Besides, to more clearly grasp the feature of light-manipulated Weyl nodes, we show their distributions with a light intensity with $eA_z/\hbar =$ 0.00, 0.06, and 0.08 \AA$^{-1}$ as illustrated in Figs. \ref{FIG3}(b), \ref{FIG3}(c), and \ref{FIG3}(d), respectively. It is found that the light irradiation forces that the Weyl nodes are away from the high-symmetry paths $\Gamma-X$ and $Z-A$ but lie in the $k_x$-$k_z$ with $k_y=0$ plane. This indicates that the incident LPL may only break the rotational symmetry $C_{2x}$ but preserve the antiunitary mirror symmetry $\tilde{M}_y$.

Next, to better understand light-controllable Weyl nodes in the light-irradiated fco-C$_6$, we perform an extensive theoretical analysis below using the low-energy effective model. Based on the crystalline symmetry of fco-C$_6$, the Weyl nodes in the absence of light fields can be described by a $2\times 2$ $k\cdot p$ Hamiltonian as
\begin{equation}\label{Hamillow}
H(\mathbf{k} )=f_x(\mathbf{k})\sigma _{x} +f_y(\mathbf{k}) \sigma _{y}+f_z(\mathbf{k}) \sigma _{z}
\end{equation}
with
\begin{equation}\label{Hamillow}
\begin{split}
&f_x(\mathbf{k})=B_{xx}(k_{x} \pm k_{x_{0} }),\\
&f_y(\mathbf{k})=B_{yy} k_{y},\\
&f_z(\mathbf{k})=B_{zx}(k_{x} \pm k_{x_{0} })^{2} +B_{zy}k_{y}^{2} +B_{zz}k_{z}(k_{z}-\frac{2\pi }{c} ),
\end{split}
\end{equation}
where we ignore the kinetic term for simplicity, $\sigma_i$ ($i=x, y, z$) are Pauli matrics, and the $k\cdot p$ parameters ${B_{ij}}$ are dependent on the crystal structure of fco-C$_6$ and can be calculated from first-principles calculations. The positions of Weyl nodes can be solved from the zero modes of the two-band Hamiltonian Eq. (\ref{Hamillow}), i.e., ($\pm k_{x_{0}}$, $0$, $0$) and ($\pm k_{x_{0}}$, $0$, $2\pi/c$). Therefore, Eq. (\ref{Hamillow}) can effectively grasp the low-energy Weyl physics of fco-C$_6$.

\begin{figure}
    \centering
    \includegraphics[width=\linewidth]{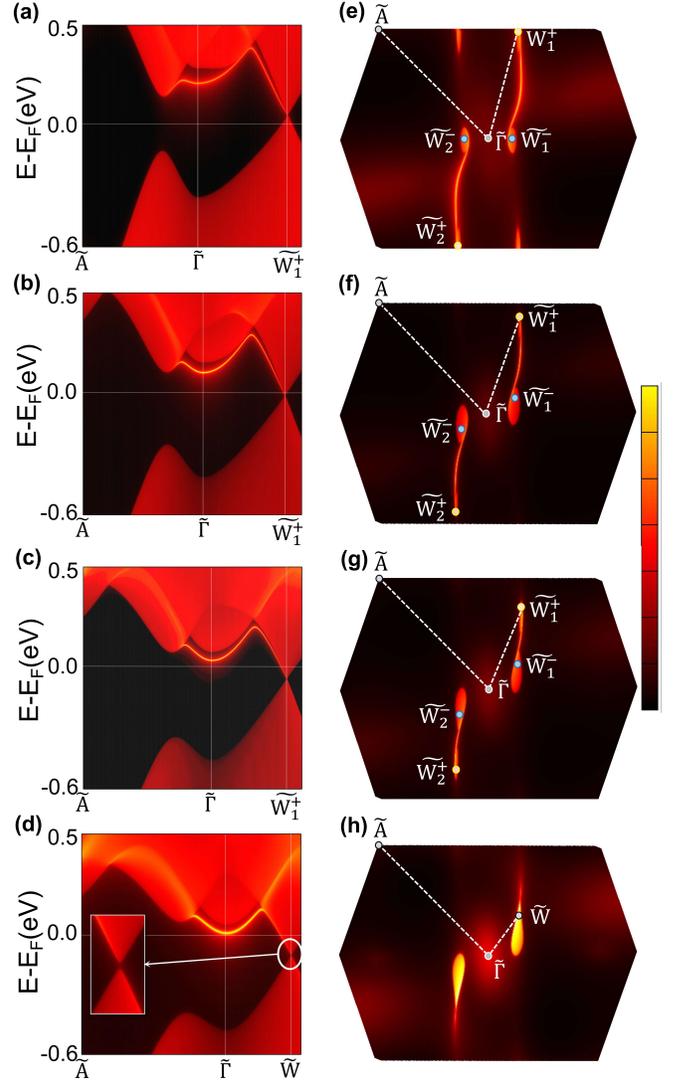}
   \caption{The calculated surface states projected on the semi-infinite (010) surface of fco-C$_{6}$ under light irradiation. Panels (a)-(d) show the LDOS with a light intensity  $eA_z/\hbar =$ 0.00, 0.06, 0.08, and 0.09 \AA$^{-1}$, respectively. Panels (e)-(h) plot the corresponding Fermi surfaces of (010)-surface BZ.
    \label{FIG4}}
\end{figure}

Then, we can obtain the photon-dressed Floquet Hamiltonian in the high frequency limit using the Floquet theorem \cite{Goldman,Eckardt2015,Bukov2015}. The effective Hamiltonian under irradiation of LPL with polarization along the $z$-axis, i.e., $\mathbf{A}(\tau)=[0, 0, A_z\sin(\omega \tau)]$, can be expressed as
\begin{equation}\label{Hamillowf}
H_F(\mathbf{k})=f_x(\mathbf{k})\sigma _{x} +f_y(\mathbf{k}) \sigma _{y}+\big[f_z(\mathbf{k})+\frac{1}{2}B_{zz}\big(\frac{eA_z}{\hbar}\big)^2\big] \sigma _{z}.
\end{equation}
Comparing Eq. (\ref{Hamillow}) and Eq. (\ref{Hamillowf}), we can see that the light irradiation of LPL introduces an extra light-corrected term $\frac{1}{2}B_{zz}\big(\frac{eA_z}{\hbar}\big)^2$ into $f_z(\mathbf{k})$. This new term will obviously change the positions of Weyl nodes. %However, the light-induced correction is momentum-independent, which implies that light irradiation of LPL does not lower the symmetry of the initial Hamiltonian Eq.(\ref{Hamillow}).
Moreover, from Eq. (\ref{Hamillowf}), we can easily obtain $C_{2x} H_F(\mathbf{k}) C_{2x}^{-1} \neq H_F(\mathbf{-k})$ and $\tilde{M}_y H_F(\mathbf{k}) \tilde{M}_y^{-1} = -H_F(\mathbf{-k})$. Thus, although the breaking of $C_{2x}$ forces  Weyl nodes away from the $\Gamma-X$ and $Z-A$ paths, the antiunitary mirror symmetry $\tilde{M}_y$ preserves the light-manipulated Weyl nodes located in the high-symmetry $k_x$-$k_z$ with $k_y=0$ plane, consistent with the results obtained from first-principles calculations. Through solving the zero-energy modes of Eq. (\ref{Hamillowf}), the positions of Weyl nodes are ($+k_x{_{0}}, $0$, k_{z_{1},{\pm}}$) and ($-k_x{_{0}}, $0$, k_{z_{2},{\pm}}$), with
\begin{equation}\label{WNpf}
\begin{split}
&k_{z_{1},{\pm}}=\frac{\pi}{c} \pm \frac{\pi}{c}\sqrt{1-\frac{1}{2}\big(\frac{eA_z}{\hbar}\big)^2 \big(\frac{c}{\pi}\big)^2},\\
&k_{z_{2},{\pm}}=-\frac{\pi}{c} \pm \frac{\pi}{c}\sqrt{1-\frac{1}{2}\big(\frac{eA_z}{\hbar}\big)^2 \big(\frac{c}{\pi}\big)^2}.
\end{split}
\end{equation}
Equation (\ref{WNpf}) indicates that the light irradiation only changes the $k_z$-component of the coordinates of Weyl nodes, which is solely dependent on the light intensity. The results obtained from the $k\cdot p$ model agree our symmetry argument. It is worth noting that the changes of $k_x$-component of Weyl nodes do not appear in Eq. (\ref{WNpf}) as the $k\cdot p$ model can only reveal the Weyl physics in the low-energy region. For the limit of weak light intensity, we can expand Eq. (\ref{WNpf}) and remain the lowest-order term. In this case, the Weyl nodes with the chirality $C=+1$ are approximately positioned at ($k_{x_{0}}$, $0$, $0$, $\delta$) and ($-k_{x_{0}}$, $0$, $0$, $-\delta$), and those with the chirality $C=-1$ are approximately positioned at ($k_{x_{0}}$, $0$, $\frac{2\pi}{c}-\delta$) and ($-k_{x_{0}}$, $0$, $-\frac{2\pi}{c}+\delta$), with $\delta=\frac{1}{4}\big(\frac{eA_z}{\hbar}\big)^2\frac{c}{\pi}$.

The Weyl physics of fco-C$_{6}$ under the irradiation of LPL can exhibit nontrivial light-manipulated surface states. To obtain the surface states, we employ the Floquet TB Hamiltonian Eq. (\ref{Hfirstf}). By employing iterative Green's function method \cite{RGFM}, we obtain the photon-dressed local density of states (LDOS) and Fermi arcs of fco-C$_{6}$. Since the two pairs of Weyl points are always located in the $k_{x}-k_{z}$ plane with $k_y = 0$ in the momentum space, we focus on studying the surface states projected on the semi-infinite (010) surface. To clearly exhibit the nontrivial properties, we depict the surface band structures pass one projected Weyl node $\tilde{W}_1^{+}$. As shown in Figs. \ref{FIG4}(a)-\ref{FIG4}(c), we plot the LDOS with a light intensity  $eA_z/\hbar =$ 0.00, 0.06, and 0.08 \AA$^{-1}$, respectively. It is found that the surface states are visibly terminated at the projections of Weyl nodes. The two projected Weyl nodes with same chirality are centrosymmetric with respect to $\tilde{\Gamma}$ as the irradiation of LPL preserves the time-reversal symmetry. The Fermi arcs originated from the nontrivial surface states crossing the Fermi level connect two projected Weyl nodes with opposite chirality [see Figs. \ref{FIG4}(e)-\ref{FIG4}(g)]. It is worth noting that the length of Fermi arcs can be controlled by light intensities. When the light intensity exceeds a threshold value, such as $eA_z/\hbar =$ 0.09 \AA$^{-1}$ in Figs. \ref{FIG4}(d) and \ref{FIG4}(h), we can see that the the projected band structure is gapped and the nontrivial Fermi arcs vanish.

\section{Summary}
In summary, by employing first-principles calculations, MLWF-based TB Hamiltonian, and Floquet theorem, we propose an approach to investigate the light-manipulated evolution of Weyl physics in realistic Weyl semimetallic materials. Under the irradiation of LPL, we reveal the positions of Weyl nodes and Fermi arcs as a function of light intensity in an ideal Weyl semimetal (i.e., carbon allotrope fco-C$_6$). Moreover, we employ a low-energy effective $k\cdot p$ model to understand light-controllable Weyl physics. The results indicate that the light-preserved symmetry allows the light-manipulated Weyl nodes can only move in the high-symmetry plane in momentum space. Our work demonstrates employing periodic driving light fields is an efficient approach to manipulate Weyl physics. Considering the carbon allotrope fco-C$_6$ with the extremely weak SOC effect, our work will draw widespread attention for light-manipulated topological states in experiments.
%\begin{figure}
%    \centering
%    \includegraphics[scale=1]{Fig1.pdf}
%    \caption{(a) Band diagram illustrating valley-dependent band inversion via hybridization of Floquet-Bloch bands (i.e., $E_c^{F}$ and $E_v^{F}$) originated from irradiation of CPL $\mathbf{A}(t)$. %(b) Schematic diagram of light-induced chiral edge channels of high Chern number VP-QAHE. The insets give the top view of lattice structures of 2D $M$Si$_2$Z$_4$ family materials and hexagonal Brillouin zone (BZ). (c) The orbital-resolved electronic band structure of monolayer VSi$_2$N$_4$ including SOC with the magnetization along the $x$-axis. The components of V $d_{xy}+d_{x^{2}-y^{2}}$ and V $d_{z^2}$ orbitals are proportional to the width of the magenta and orange lines, respectively. (d) The photon-dressed band structures of VSi$_2$N$_4$ subject to left-handed CPL with a certain light intensity and frequency (i.e., $\hbar \omega=0.145$ eV and $eA/\hbar=0.028$ {\AA}$^{-1}$). The black solid lines represent the equilibrium bands. The blue and red dashed lines represent Floquet-Bloch bands created by absorbtion and emission of photons, respectively. The insets indicate that two Floquet-Bloch bands invert at the $K$ point and preserve the trivial band gap at the $K'$ point.
%    \label{FIG1}}
%\end{figure}

%in experimentally accessible regimes

\section{Acknowledgments}
This work was supported by the National Natural Science Foundation of China (NSFC, Grants No.~12247181, No.~12222402, No. 12074108, No.~11974062, and No.~12147102) and the Natural Science Foundation of Chongqing (Grants No. CSTB2022NSCQ-MSX0568 and No. CSTB2023NSCQ-JQX0024), and the Fundamental Research Funds for the Central Universities (Grant No. 2023CDJXY-048). Simulations were performed on Hefei advanced computing center.

%\bibliographystyle{apsrev4-2}
%\bibliography{BCT-Floquet-ref}

\begin{thebibliography}{61}%
\makeatletter
\providecommand \@ifxundefined [1]{%
 \@ifx{#1\undefined}
}%
\providecommand \@ifnum [1]{%
 \ifnum #1\expandafter \@firstoftwo
 \else \expandafter \@secondoftwo
 \fi
}%
\providecommand \@ifx [1]{%
 \ifx #1\expandafter \@firstoftwo
 \else \expandafter \@secondoftwo
 \fi
}%
\providecommand \natexlab [1]{#1}%
\providecommand \enquote  [1]{``#1''}%
\providecommand \bibnamefont  [1]{#1}%
\providecommand \bibfnamefont [1]{#1}%
\providecommand \citenamefont [1]{#1}%
\providecommand \href@noop [0]{\@secondoftwo}%
\providecommand \href [0]{\begingroup \@sanitize@url \@href}%
\providecommand \@href[1]{\@@startlink{#1}\@@href}%
\providecommand \@@href[1]{\endgroup#1\@@endlink}%
\providecommand \@sanitize@url [0]{\catcode `\\12\catcode `\$12\catcode
  `\&12\catcode `\#12\catcode `\^12\catcode `\_12\catcode `\%12\relax}%
\providecommand \@@startlink[1]{}%
\providecommand \@@endlink[0]{}%
\providecommand \url  [0]{\begingroup\@sanitize@url \@url }%
\providecommand \@url [1]{\endgroup\@href {#1}{\urlprefix }}%
\providecommand \urlprefix  [0]{URL }%
\providecommand \Eprint [0]{\href }%
\providecommand \doibase [0]{https://doi.org/}%
\providecommand \selectlanguage [0]{\@gobble}%
\providecommand \bibinfo  [0]{\@secondoftwo}%
\providecommand \bibfield  [0]{\@secondoftwo}%
\providecommand \translation [1]{[#1]}%
\providecommand \BibitemOpen [0]{}%
\providecommand \bibitemStop [0]{}%
\providecommand \bibitemNoStop [0]{.\EOS\space}%
\providecommand \EOS [0]{\spacefactor3000\relax}%
\providecommand \BibitemShut  [1]{\csname bibitem#1\endcsname}%
\let\auto@bib@innerbib\@empty
%</preamble>
\bibitem [{\citenamefont {Hasan}\ and\ \citenamefont
  {Kane}(2010)}]{Kane-RevModPhys.82.3045}%
  \BibitemOpen
  \bibfield  {author} {\bibinfo {author} {\bibfnamefont {M.~Z.}\ \bibnamefont
  {Hasan}}\ and\ \bibinfo {author} {\bibfnamefont {C.~L.}\ \bibnamefont
  {Kane}},\ }\href {https://doi.org/10.1103/RevModPhys.82.3045} {\bibfield
  {journal} {\bibinfo  {journal} {Rev. Mod. Phys.}\ }\textbf {\bibinfo {volume}
  {82}},\ \bibinfo {pages} {3045} (\bibinfo {year} {2010})}\BibitemShut
  {NoStop}%
\bibitem [{\citenamefont {Qi}\ and\ \citenamefont
  {Zhang}(2011)}]{ZSC-RevModPhys.83.1057}%
  \BibitemOpen
  \bibfield  {author} {\bibinfo {author} {\bibfnamefont {X.-L.}\ \bibnamefont
  {Qi}}\ and\ \bibinfo {author} {\bibfnamefont {S.-C.}\ \bibnamefont {Zhang}},\
  }\href {https://doi.org/10.1103/RevModPhys.83.1057} {\bibfield  {journal}
  {\bibinfo  {journal} {Rev. Mod. Phys.}\ }\textbf {\bibinfo {volume} {83}},\
  \bibinfo {pages} {1057} (\bibinfo {year} {2011})}\BibitemShut {NoStop}%
\bibitem [{\citenamefont {Armitage}\ \emph {et~al.}(2018)\citenamefont
  {Armitage}, \citenamefont {Mele},\ and\ \citenamefont
  {Vishwanath}}]{RevModPhys.90.015001}%
  \BibitemOpen
  \bibfield  {author} {\bibinfo {author} {\bibfnamefont {N.~P.}\ \bibnamefont
  {Armitage}}, \bibinfo {author} {\bibfnamefont {E.~J.}\ \bibnamefont {Mele}},\
  and\ \bibinfo {author} {\bibfnamefont {A.}~\bibnamefont {Vishwanath}},\
  }\href {https://doi.org/10.1103/RevModPhys.90.015001} {\bibfield  {journal}
  {\bibinfo  {journal} {Rev. Mod. Phys.}\ }\textbf {\bibinfo {volume} {90}},\
  \bibinfo {pages} {015001} (\bibinfo {year} {2018})}\BibitemShut {NoStop}%
\bibitem [{\citenamefont {Wang}\ \emph {et~al.}(2012)\citenamefont {Wang},
  \citenamefont {Sun}, \citenamefont {Chen}, \citenamefont {Franchini},
  \citenamefont {Xu}, \citenamefont {Weng}, \citenamefont {Dai},\ and\
  \citenamefont {Fang}}]{WangPhysRevB.85.195320}%
  \BibitemOpen
  \bibfield  {author} {\bibinfo {author} {\bibfnamefont {Z.}~\bibnamefont
  {Wang}}, \bibinfo {author} {\bibfnamefont {Y.}~\bibnamefont {Sun}}, \bibinfo
  {author} {\bibfnamefont {X.-Q.}\ \bibnamefont {Chen}}, \bibinfo {author}
  {\bibfnamefont {C.}~\bibnamefont {Franchini}}, \bibinfo {author}
  {\bibfnamefont {G.}~\bibnamefont {Xu}}, \bibinfo {author} {\bibfnamefont
  {H.}~\bibnamefont {Weng}}, \bibinfo {author} {\bibfnamefont {X.}~\bibnamefont
  {Dai}},\ and\ \bibinfo {author} {\bibfnamefont {Z.}~\bibnamefont {Fang}},\
  }\href {https://doi.org/10.1103/PhysRevB.85.195320} {\bibfield  {journal}
  {\bibinfo  {journal} {Phys. Rev. B}\ }\textbf {\bibinfo {volume} {85}},\
  \bibinfo {pages} {195320} (\bibinfo {year} {2012})}\BibitemShut {NoStop}%
\bibitem [{\citenamefont {Liu}\ \emph {et~al.}(2014)\citenamefont {Liu},
  \citenamefont {Zhou}, \citenamefont {Zhang}, \citenamefont {Wang},
  \citenamefont {Weng}, \citenamefont {Prabhakaran}, \citenamefont {Mo},
  \citenamefont {Shen}, \citenamefont {Fang}, \citenamefont {Dai},
  \citenamefont {Hussain},\ and\ \citenamefont {Chen}}]{ScienceLiu864}%
  \BibitemOpen
  \bibfield  {author} {\bibinfo {author} {\bibfnamefont {Z.~K.}\ \bibnamefont
  {Liu}}, \bibinfo {author} {\bibfnamefont {B.}~\bibnamefont {Zhou}}, \bibinfo
  {author} {\bibfnamefont {Y.}~\bibnamefont {Zhang}}, \bibinfo {author}
  {\bibfnamefont {Z.~J.}\ \bibnamefont {Wang}}, \bibinfo {author}
  {\bibfnamefont {H.~M.}\ \bibnamefont {Weng}}, \bibinfo {author}
  {\bibfnamefont {D.}~\bibnamefont {Prabhakaran}}, \bibinfo {author}
  {\bibfnamefont {S.-K.}\ \bibnamefont {Mo}}, \bibinfo {author} {\bibfnamefont
  {Z.~X.}\ \bibnamefont {Shen}}, \bibinfo {author} {\bibfnamefont
  {Z.}~\bibnamefont {Fang}}, \bibinfo {author} {\bibfnamefont {X.}~\bibnamefont
  {Dai}}, \bibinfo {author} {\bibfnamefont {Z.}~\bibnamefont {Hussain}},\ and\
  \bibinfo {author} {\bibfnamefont {Y.~L.}\ \bibnamefont {Chen}},\ }\href
  {https://doi.org/10.1126/science.1245085} {\bibfield  {journal} {\bibinfo
  {journal} {Science}\ }\textbf {\bibinfo {volume} {343}},\ \bibinfo {pages}
  {864} (\bibinfo {year} {2014})}\BibitemShut {NoStop}%
\bibitem [{\citenamefont {Wan}\ \emph {et~al.}(2011)\citenamefont {Wan},
  \citenamefont {Turner}, \citenamefont {Vishwanath},\ and\ \citenamefont
  {Savrasov}}]{Wan2011}%
  \BibitemOpen
  \bibfield  {author} {\bibinfo {author} {\bibfnamefont {X.}~\bibnamefont
  {Wan}}, \bibinfo {author} {\bibfnamefont {A.~M.}\ \bibnamefont {Turner}},
  \bibinfo {author} {\bibfnamefont {A.}~\bibnamefont {Vishwanath}},\ and\
  \bibinfo {author} {\bibfnamefont {S.~Y.}\ \bibnamefont {Savrasov}},\ }\href
  {https://doi.org/10.1103/PhysRevB.83.205101} {\bibfield  {journal} {\bibinfo
  {journal} {Phys. Rev. B}\ }\textbf {\bibinfo {volume} {83}},\ \bibinfo
  {pages} {205101} (\bibinfo {year} {2011})}\BibitemShut {NoStop}%
\bibitem [{\citenamefont {Xu}\ \emph {et~al.}(2011)\citenamefont {Xu},
  \citenamefont {Weng}, \citenamefont {Wang}, \citenamefont {Dai},\ and\
  \citenamefont {Fang}}]{Xu2011}%
  \BibitemOpen
  \bibfield  {author} {\bibinfo {author} {\bibfnamefont {G.}~\bibnamefont
  {Xu}}, \bibinfo {author} {\bibfnamefont {H.}~\bibnamefont {Weng}}, \bibinfo
  {author} {\bibfnamefont {Z.}~\bibnamefont {Wang}}, \bibinfo {author}
  {\bibfnamefont {X.}~\bibnamefont {Dai}},\ and\ \bibinfo {author}
  {\bibfnamefont {Z.}~\bibnamefont {Fang}},\ }\href
  {https://doi.org/10.1103/PhysRevLett.107.186806} {\bibfield  {journal}
  {\bibinfo  {journal} {Phys. Rev. Lett.}\ }\textbf {\bibinfo {volume} {107}},\
  \bibinfo {pages} {186806} (\bibinfo {year} {2011})}\BibitemShut {NoStop}%
\bibitem [{\citenamefont {Weng}\ \emph {et~al.}(2015)\citenamefont {Weng},
  \citenamefont {Fang}, \citenamefont {Fang}, \citenamefont {Bernevig},\ and\
  \citenamefont {Dai}}]{PhysRevX.5.011029}%
  \BibitemOpen
  \bibfield  {author} {\bibinfo {author} {\bibfnamefont {H.}~\bibnamefont
  {Weng}}, \bibinfo {author} {\bibfnamefont {C.}~\bibnamefont {Fang}}, \bibinfo
  {author} {\bibfnamefont {Z.}~\bibnamefont {Fang}}, \bibinfo {author}
  {\bibfnamefont {B.~A.}\ \bibnamefont {Bernevig}},\ and\ \bibinfo {author}
  {\bibfnamefont {X.}~\bibnamefont {Dai}},\ }\href
  {https://doi.org/10.1103/PhysRevX.5.011029} {\bibfield  {journal} {\bibinfo
  {journal} {Phys. Rev. X}\ }\textbf {\bibinfo {volume} {5}},\ \bibinfo {pages}
  {011029} (\bibinfo {year} {2015})}\BibitemShut {NoStop}%
\bibitem [{\citenamefont {Xu}\ \emph {et~al.}(2015)\citenamefont {Xu},
  \citenamefont {Belopolski}, \citenamefont {Alidoust}, \citenamefont
  {Neupane}, \citenamefont {Bian}, \citenamefont {Zhang}, \citenamefont
  {Sankar}, \citenamefont {Chang}, \citenamefont {Yuan}, \citenamefont {Lee},
  \citenamefont {Huang}, \citenamefont {Zheng}, \citenamefont {Ma},
  \citenamefont {Sanchez}, \citenamefont {Wang}, \citenamefont {Bansil},
  \citenamefont {Chou}, \citenamefont {Shibayev}, \citenamefont {Lin},
  \citenamefont {Jia},\ and\ \citenamefont {Hasan}}]{Xu2015}%
  \BibitemOpen
  \bibfield  {author} {\bibinfo {author} {\bibfnamefont {S.-Y.}\ \bibnamefont
  {Xu}}, \bibinfo {author} {\bibfnamefont {I.}~\bibnamefont {Belopolski}},
  \bibinfo {author} {\bibfnamefont {N.}~\bibnamefont {Alidoust}}, \bibinfo
  {author} {\bibfnamefont {M.}~\bibnamefont {Neupane}}, \bibinfo {author}
  {\bibfnamefont {G.}~\bibnamefont {Bian}}, \bibinfo {author} {\bibfnamefont
  {C.}~\bibnamefont {Zhang}}, \bibinfo {author} {\bibfnamefont
  {R.}~\bibnamefont {Sankar}}, \bibinfo {author} {\bibfnamefont
  {G.}~\bibnamefont {Chang}}, \bibinfo {author} {\bibfnamefont
  {Z.}~\bibnamefont {Yuan}}, \bibinfo {author} {\bibfnamefont {C.-C.}\
  \bibnamefont {Lee}}, \bibinfo {author} {\bibfnamefont {S.-M.}\ \bibnamefont
  {Huang}}, \bibinfo {author} {\bibfnamefont {H.}~\bibnamefont {Zheng}},
  \bibinfo {author} {\bibfnamefont {J.}~\bibnamefont {Ma}}, \bibinfo {author}
  {\bibfnamefont {D.~S.}\ \bibnamefont {Sanchez}}, \bibinfo {author}
  {\bibfnamefont {B.}~\bibnamefont {Wang}}, \bibinfo {author} {\bibfnamefont
  {A.}~\bibnamefont {Bansil}}, \bibinfo {author} {\bibfnamefont
  {F.}~\bibnamefont {Chou}}, \bibinfo {author} {\bibfnamefont {P.~P.}\
  \bibnamefont {Shibayev}}, \bibinfo {author} {\bibfnamefont {H.}~\bibnamefont
  {Lin}}, \bibinfo {author} {\bibfnamefont {S.}~\bibnamefont {Jia}},\ and\
  \bibinfo {author} {\bibfnamefont {M.~Z.}\ \bibnamefont {Hasan}},\ }\href
  {https://doi.org/10.1126/science.aaa9297} {\bibfield  {journal} {\bibinfo
  {journal} {Science}\ }\textbf {\bibinfo {volume} {349}},\ \bibinfo {pages}
  {613} (\bibinfo {year} {2015})}\BibitemShut {NoStop}%
\bibitem [{\citenamefont {Burkov}\ \emph {et~al.}(2011)\citenamefont {Burkov},
  \citenamefont {Hook},\ and\ \citenamefont {Balents}}]{PhysRevB.84.235126}%
  \BibitemOpen
  \bibfield  {author} {\bibinfo {author} {\bibfnamefont {A.~A.}\ \bibnamefont
  {Burkov}}, \bibinfo {author} {\bibfnamefont {M.~D.}\ \bibnamefont {Hook}},\
  and\ \bibinfo {author} {\bibfnamefont {L.}~\bibnamefont {Balents}},\ }\href
  {https://doi.org/10.1103/PhysRevB.84.235126} {\bibfield  {journal} {\bibinfo
  {journal} {Phys. Rev. B}\ }\textbf {\bibinfo {volume} {84}},\ \bibinfo
  {pages} {235126} (\bibinfo {year} {2011})}\BibitemShut {NoStop}%
\bibitem [{\citenamefont {Kim}\ \emph {et~al.}(2015)\citenamefont {Kim},
  \citenamefont {Wieder}, \citenamefont {Kane},\ and\ \citenamefont
  {Rappe}}]{PhysRevLett.115.036806}%
  \BibitemOpen
  \bibfield  {author} {\bibinfo {author} {\bibfnamefont {Y.}~\bibnamefont
  {Kim}}, \bibinfo {author} {\bibfnamefont {B.~J.}\ \bibnamefont {Wieder}},
  \bibinfo {author} {\bibfnamefont {C.~L.}\ \bibnamefont {Kane}},\ and\
  \bibinfo {author} {\bibfnamefont {A.~M.}\ \bibnamefont {Rappe}},\ }\href
  {https://doi.org/10.1103/PhysRevLett.115.036806} {\bibfield  {journal}
  {\bibinfo  {journal} {Phys. Rev. Lett.}\ }\textbf {\bibinfo {volume} {115}},\
  \bibinfo {pages} {036806} (\bibinfo {year} {2015})}\BibitemShut {NoStop}%
\bibitem [{\citenamefont {Yu}\ \emph {et~al.}(2015)\citenamefont {Yu},
  \citenamefont {Weng}, \citenamefont {Fang}, \citenamefont {Dai},\ and\
  \citenamefont {Hu}}]{PhysRevLett.115.036807}%
  \BibitemOpen
  \bibfield  {author} {\bibinfo {author} {\bibfnamefont {R.}~\bibnamefont
  {Yu}}, \bibinfo {author} {\bibfnamefont {H.}~\bibnamefont {Weng}}, \bibinfo
  {author} {\bibfnamefont {Z.}~\bibnamefont {Fang}}, \bibinfo {author}
  {\bibfnamefont {X.}~\bibnamefont {Dai}},\ and\ \bibinfo {author}
  {\bibfnamefont {X.}~\bibnamefont {Hu}},\ }\href
  {https://doi.org/10.1103/PhysRevLett.115.036807} {\bibfield  {journal}
  {\bibinfo  {journal} {Phys. Rev. Lett.}\ }\textbf {\bibinfo {volume} {115}},\
  \bibinfo {pages} {036807} (\bibinfo {year} {2015})}\BibitemShut {NoStop}%
\bibitem [{\citenamefont {Soluyanov}\ \emph {et~al.}(2015)\citenamefont
  {Soluyanov}, \citenamefont {Gresch}, \citenamefont {Wang}, \citenamefont
  {Wu},\ and\ \citenamefont {Troyer}}]{NatureSoluyanov2015}%
  \BibitemOpen
  \bibfield  {author} {\bibinfo {author} {\bibfnamefont {A.~A.}\ \bibnamefont
  {Soluyanov}}, \bibinfo {author} {\bibfnamefont {D.}~\bibnamefont {Gresch}},
  \bibinfo {author} {\bibfnamefont {Z.}~\bibnamefont {Wang}}, \bibinfo {author}
  {\bibfnamefont {Q.}~\bibnamefont {Wu}},\ and\ \bibinfo {author}
  {\bibfnamefont {M.}~\bibnamefont {Troyer}},\ }\href
  {https://doi.org/https://doi.org/10.1038/nature15768} {\bibfield  {journal}
  {\bibinfo  {journal} {Nature}\ }\textbf {\bibinfo {volume} {527}},\ \bibinfo
  {pages} {495} (\bibinfo {year} {2015})}\BibitemShut {NoStop}%
\bibitem [{\citenamefont {Zhu}\ \emph {et~al.}(2016)\citenamefont {Zhu},
  \citenamefont {Winkler}, \citenamefont {Wu}, \citenamefont {Li},\ and\
  \citenamefont {Soluyanov}}]{ZhuPhysRevX.6.031003}%
  \BibitemOpen
  \bibfield  {author} {\bibinfo {author} {\bibfnamefont {Z.}~\bibnamefont
  {Zhu}}, \bibinfo {author} {\bibfnamefont {G.~W.}\ \bibnamefont {Winkler}},
  \bibinfo {author} {\bibfnamefont {Q.}~\bibnamefont {Wu}}, \bibinfo {author}
  {\bibfnamefont {J.}~\bibnamefont {Li}},\ and\ \bibinfo {author}
  {\bibfnamefont {A.~A.}\ \bibnamefont {Soluyanov}},\ }\href
  {https://doi.org/10.1103/PhysRevX.6.031003} {\bibfield  {journal} {\bibinfo
  {journal} {Phys. Rev. X}\ }\textbf {\bibinfo {volume} {6}},\ \bibinfo {pages}
  {031003} (\bibinfo {year} {2016})}\BibitemShut {NoStop}%
\bibitem [{\citenamefont {Lv}\ \emph {et~al.}(2017)\citenamefont {Lv},
  \citenamefont {Feng}, \citenamefont {Xu}, \citenamefont {Gao}, \citenamefont
  {Ma}, \citenamefont {Kong}, \citenamefont {Richard}, \citenamefont {Huang},
  \citenamefont {Strocov}, \citenamefont {Fang}, \citenamefont {Weng},
  \citenamefont {Shi}, \citenamefont {Tian},\ and\ \citenamefont
  {H.}}]{lvNature2017}%
  \BibitemOpen
  \bibfield  {author} {\bibinfo {author} {\bibfnamefont {B.~Q.}\ \bibnamefont
  {Lv}}, \bibinfo {author} {\bibfnamefont {Z.-L.}\ \bibnamefont {Feng}},
  \bibinfo {author} {\bibfnamefont {Q.-N.}\ \bibnamefont {Xu}}, \bibinfo
  {author} {\bibfnamefont {X.}~\bibnamefont {Gao}}, \bibinfo {author}
  {\bibfnamefont {J.-Z.}\ \bibnamefont {Ma}}, \bibinfo {author} {\bibfnamefont
  {L.-Y.}\ \bibnamefont {Kong}}, \bibinfo {author} {\bibfnamefont
  {P.}~\bibnamefont {Richard}}, \bibinfo {author} {\bibfnamefont {Y.-B.}\
  \bibnamefont {Huang}}, \bibinfo {author} {\bibfnamefont {V.~N.}\ \bibnamefont
  {Strocov}}, \bibinfo {author} {\bibfnamefont {C.}~\bibnamefont {Fang}},
  \bibinfo {author} {\bibfnamefont {H.-M.}\ \bibnamefont {Weng}}, \bibinfo
  {author} {\bibfnamefont {Y.-G.}\ \bibnamefont {Shi}}, \bibinfo {author}
  {\bibfnamefont {T.~D.}\ \bibnamefont {Tian}},\ and\ \bibinfo {author}
  {\bibnamefont {H.}},\ }\href
  {https://doi.org/https://doi.org/10.1038/nature22390} {\bibfield  {journal}
  {\bibinfo  {journal} {Nature}\ }\textbf {\bibinfo {volume} {546}},\ \bibinfo
  {pages} {627} (\bibinfo {year} {2017})}\BibitemShut {NoStop}%
\bibitem [{\citenamefont {Wang}\ \emph {et~al.}(2016)\citenamefont {Wang},
  \citenamefont {Alexandradinata}, \citenamefont {Cava},\ and\ \citenamefont
  {Bernevig}}]{Wang2016}%
  \BibitemOpen
  \bibfield  {author} {\bibinfo {author} {\bibfnamefont {Z.}~\bibnamefont
  {Wang}}, \bibinfo {author} {\bibfnamefont {A.}~\bibnamefont
  {Alexandradinata}}, \bibinfo {author} {\bibfnamefont {R.~J.}\ \bibnamefont
  {Cava}},\ and\ \bibinfo {author} {\bibfnamefont {B.~A.}\ \bibnamefont
  {Bernevig}},\ }\href {https://doi.org/10.1038/nature17410} {\bibfield
  {journal} {\bibinfo  {journal} {Nature}\ }\textbf {\bibinfo {volume} {532}},\
  \bibinfo {pages} {189} (\bibinfo {year} {2016})}\BibitemShut {NoStop}%
\bibitem [{\citenamefont {Wang}\ \emph {et~al.}(2019)\citenamefont {Wang},
  \citenamefont {Liu},\ and\ \citenamefont {Zhu}}]{PhysRevLett.123.126403}%
  \BibitemOpen
  \bibfield  {author} {\bibinfo {author} {\bibfnamefont {Z.~F.}\ \bibnamefont
  {Wang}}, \bibinfo {author} {\bibfnamefont {B.}~\bibnamefont {Liu}},\ and\
  \bibinfo {author} {\bibfnamefont {W.}~\bibnamefont {Zhu}},\ }\href
  {https://doi.org/10.1103/PhysRevLett.123.126403} {\bibfield  {journal}
  {\bibinfo  {journal} {Phys. Rev. Lett.}\ }\textbf {\bibinfo {volume} {123}},\
  \bibinfo {pages} {126403} (\bibinfo {year} {2019})}\BibitemShut {NoStop}%
\bibitem [{\citenamefont {Wang}\ \emph
  {et~al.}(2017{\natexlab{a}})\citenamefont {Wang}, \citenamefont {Liu},
  \citenamefont {Yu}, \citenamefont {Sheng},\ and\ \citenamefont
  {Yang}}]{Hourglass}%
  \BibitemOpen
  \bibfield  {author} {\bibinfo {author} {\bibfnamefont {S.-S.}\ \bibnamefont
  {Wang}}, \bibinfo {author} {\bibfnamefont {Y.}~\bibnamefont {Liu}}, \bibinfo
  {author} {\bibfnamefont {Z.-M.}\ \bibnamefont {Yu}}, \bibinfo {author}
  {\bibfnamefont {X.-L.}\ \bibnamefont {Sheng}},\ and\ \bibinfo {author}
  {\bibfnamefont {S.~A.}\ \bibnamefont {Yang}},\ }\href
  {https://doi.org/10.1038/s41467-017-01986-3} {\bibfield  {journal} {\bibinfo
  {journal} {Nat. Commun.}\ }\textbf {\bibinfo {volume} {8}},\ \bibinfo {pages}
  {1844} (\bibinfo {year} {2017}{\natexlab{a}})}\BibitemShut {NoStop}%
\bibitem [{\citenamefont {Bradlyn}\ \emph {et~al.}(2016)\citenamefont
  {Bradlyn}, \citenamefont {Cano}, \citenamefont {Wang}, \citenamefont
  {Vergniory}, \citenamefont {Felser}, \citenamefont {Cava},\ and\
  \citenamefont {Bernevig}}]{Bradlynaaf5037}%
  \BibitemOpen
  \bibfield  {author} {\bibinfo {author} {\bibfnamefont {B.}~\bibnamefont
  {Bradlyn}}, \bibinfo {author} {\bibfnamefont {J.}~\bibnamefont {Cano}},
  \bibinfo {author} {\bibfnamefont {Z.}~\bibnamefont {Wang}}, \bibinfo {author}
  {\bibfnamefont {M.~G.}\ \bibnamefont {Vergniory}}, \bibinfo {author}
  {\bibfnamefont {C.}~\bibnamefont {Felser}}, \bibinfo {author} {\bibfnamefont
  {R.~J.}\ \bibnamefont {Cava}},\ and\ \bibinfo {author} {\bibfnamefont
  {B.~A.}\ \bibnamefont {Bernevig}},\ }\href
  {https://doi.org/10.1126/science.aaf5037} {\bibfield  {journal} {\bibinfo
  {journal} {Science}\ }\textbf {\bibinfo {volume} {353}},\ \bibinfo {pages}
  {6299} (\bibinfo {year} {2016})}\BibitemShut {NoStop}%
\bibitem [{\citenamefont {Fang}\ \emph {et~al.}(2003)\citenamefont {Fang},
  \citenamefont {Nagaosa}, \citenamefont {Takahashi}, \citenamefont {Asamitsu},
  \citenamefont {Mathieu}, \citenamefont {Ogasawara}, \citenamefont {Yamada},
  \citenamefont {Kawasaki}, \citenamefont {Tokura},\ and\ \citenamefont
  {Terakura}}]{Sciencemonople2003}%
  \BibitemOpen
  \bibfield  {author} {\bibinfo {author} {\bibfnamefont {Z.}~\bibnamefont
  {Fang}}, \bibinfo {author} {\bibfnamefont {N.}~\bibnamefont {Nagaosa}},
  \bibinfo {author} {\bibfnamefont {K.~S.}\ \bibnamefont {Takahashi}}, \bibinfo
  {author} {\bibfnamefont {A.}~\bibnamefont {Asamitsu}}, \bibinfo {author}
  {\bibfnamefont {R.}~\bibnamefont {Mathieu}}, \bibinfo {author} {\bibfnamefont
  {T.}~\bibnamefont {Ogasawara}}, \bibinfo {author} {\bibfnamefont
  {H.}~\bibnamefont {Yamada}}, \bibinfo {author} {\bibfnamefont
  {M.}~\bibnamefont {Kawasaki}}, \bibinfo {author} {\bibfnamefont
  {Y.}~\bibnamefont {Tokura}},\ and\ \bibinfo {author} {\bibfnamefont
  {K.}~\bibnamefont {Terakura}},\ }\href
  {https://doi.org/10.1126/science.1089408} {\bibfield  {journal} {\bibinfo
  {journal} {Science}\ }\textbf {\bibinfo {volume} {302}},\ \bibinfo {pages}
  {92} (\bibinfo {year} {2003})}\BibitemShut {NoStop}%
\bibitem [{\citenamefont {Wang}\ \emph
  {et~al.}(2017{\natexlab{b}})\citenamefont {Wang}, \citenamefont {Zhao},
  \citenamefont {Jin}, \citenamefont {Xu}, \citenamefont {Gan}, \citenamefont
  {Wu}, \citenamefont {Xu},\ and\ \citenamefont {Tong}}]{Wang2017}%
  \BibitemOpen
  \bibfield  {author} {\bibinfo {author} {\bibfnamefont {R.}~\bibnamefont
  {Wang}}, \bibinfo {author} {\bibfnamefont {J.~Z.}\ \bibnamefont {Zhao}},
  \bibinfo {author} {\bibfnamefont {Y.~J.}\ \bibnamefont {Jin}}, \bibinfo
  {author} {\bibfnamefont {W.~P.}\ \bibnamefont {Xu}}, \bibinfo {author}
  {\bibfnamefont {L.-Y.}\ \bibnamefont {Gan}}, \bibinfo {author} {\bibfnamefont
  {X.~Z.}\ \bibnamefont {Wu}}, \bibinfo {author} {\bibfnamefont
  {H.}~\bibnamefont {Xu}},\ and\ \bibinfo {author} {\bibfnamefont {S.~Y.}\
  \bibnamefont {Tong}},\ }\href {https://doi.org/10.1103/PhysRevB.96.121104}
  {\bibfield  {journal} {\bibinfo  {journal} {Phys. Rev. B}\ }\textbf {\bibinfo
  {volume} {96}},\ \bibinfo {pages} {121104} (\bibinfo {year}
  {2017}{\natexlab{b}})}\BibitemShut {NoStop}%
\bibitem [{\citenamefont {Jin}\ \emph {et~al.}(2017)\citenamefont {Jin},
  \citenamefont {Wang}, \citenamefont {Chen}, \citenamefont {Zhao},
  \citenamefont {Zhao},\ and\ \citenamefont {Xu}}]{Wang20172}%
  \BibitemOpen
  \bibfield  {author} {\bibinfo {author} {\bibfnamefont {Y.~J.}\ \bibnamefont
  {Jin}}, \bibinfo {author} {\bibfnamefont {R.}~\bibnamefont {Wang}}, \bibinfo
  {author} {\bibfnamefont {Z.~J.}\ \bibnamefont {Chen}}, \bibinfo {author}
  {\bibfnamefont {J.~Z.}\ \bibnamefont {Zhao}}, \bibinfo {author}
  {\bibfnamefont {Y.~J.}\ \bibnamefont {Zhao}},\ and\ \bibinfo {author}
  {\bibfnamefont {H.}~\bibnamefont {Xu}},\ }\href
  {https://doi.org/10.1103/PhysRevB.96.201102} {\bibfield  {journal} {\bibinfo
  {journal} {Phys. Rev. B}\ }\textbf {\bibinfo {volume} {96}},\ \bibinfo
  {pages} {201102} (\bibinfo {year} {2017})}\BibitemShut {NoStop}%
\bibitem [{\citenamefont {Xia}\ \emph {et~al.}(2019)\citenamefont {Xia},
  \citenamefont {Jin}, \citenamefont {Zhao}, \citenamefont {Chen},
  \citenamefont {Zheng}, \citenamefont {Zhao}, \citenamefont {Wang},\ and\
  \citenamefont {Xu}}]{WangPRL2019}%
  \BibitemOpen
  \bibfield  {author} {\bibinfo {author} {\bibfnamefont {B.~W.}\ \bibnamefont
  {Xia}}, \bibinfo {author} {\bibfnamefont {Y.~J.}\ \bibnamefont {Jin}},
  \bibinfo {author} {\bibfnamefont {J.~Z.}\ \bibnamefont {Zhao}}, \bibinfo
  {author} {\bibfnamefont {Z.~J.}\ \bibnamefont {Chen}}, \bibinfo {author}
  {\bibfnamefont {B.~B.}\ \bibnamefont {Zheng}}, \bibinfo {author}
  {\bibfnamefont {Y.~J.}\ \bibnamefont {Zhao}}, \bibinfo {author}
  {\bibfnamefont {R.}~\bibnamefont {Wang}},\ and\ \bibinfo {author}
  {\bibfnamefont {H.}~\bibnamefont {Xu}},\ }\href
  {https://doi.org/10.1103/PhysRevLett.122.057205} {\bibfield  {journal}
  {\bibinfo  {journal} {Phys. Rev. Lett.}\ }\textbf {\bibinfo {volume} {122}},\
  \bibinfo {pages} {057205} (\bibinfo {year} {2019})}\BibitemShut {NoStop}%
\bibitem [{\citenamefont {Morali}\ \emph {et~al.}(2019)\citenamefont {Morali},
  \citenamefont {Batabyal}, \citenamefont {Nag}, \citenamefont {Liu},
  \citenamefont {Xu}, \citenamefont {Sun}, \citenamefont {Yan}, \citenamefont
  {Felser}, \citenamefont {Avraham},\ and\ \citenamefont
  {Beidenkopf}}]{Morali1286}%
  \BibitemOpen
  \bibfield  {author} {\bibinfo {author} {\bibfnamefont {N.}~\bibnamefont
  {Morali}}, \bibinfo {author} {\bibfnamefont {R.}~\bibnamefont {Batabyal}},
  \bibinfo {author} {\bibfnamefont {P.~K.}\ \bibnamefont {Nag}}, \bibinfo
  {author} {\bibfnamefont {E.}~\bibnamefont {Liu}}, \bibinfo {author}
  {\bibfnamefont {Q.}~\bibnamefont {Xu}}, \bibinfo {author} {\bibfnamefont
  {Y.}~\bibnamefont {Sun}}, \bibinfo {author} {\bibfnamefont {B.}~\bibnamefont
  {Yan}}, \bibinfo {author} {\bibfnamefont {C.}~\bibnamefont {Felser}},
  \bibinfo {author} {\bibfnamefont {N.}~\bibnamefont {Avraham}},\ and\ \bibinfo
  {author} {\bibfnamefont {H.}~\bibnamefont {Beidenkopf}},\ }\href
  {https://doi.org/10.1126/science.aav2334} {\bibfield  {journal} {\bibinfo
  {journal} {Science}\ }\textbf {\bibinfo {volume} {365}},\ \bibinfo {pages}
  {1286} (\bibinfo {year} {2019})}\BibitemShut {NoStop}%
\bibitem [{\citenamefont {Belopolski}\ \emph {et~al.}(2019)\citenamefont
  {Belopolski}, \citenamefont {Manna}, \citenamefont {Sanchez}, \citenamefont
  {Chang}, \citenamefont {Ernst}, \citenamefont {Yin}, \citenamefont {Zhang},
  \citenamefont {Cochran}, \citenamefont {Shumiya}, \citenamefont {Zheng},
  \citenamefont {Singh}, \citenamefont {Bian}, \citenamefont {Multer},
  \citenamefont {Litskevich}, \citenamefont {Zhou}, \citenamefont {Huang},
  \citenamefont {Wang}, \citenamefont {Chang}, \citenamefont {Xu},
  \citenamefont {Bansil}, \citenamefont {Felser}, \citenamefont {Lin},\ and\
  \citenamefont {Hasan}}]{Belopolski1278}%
  \BibitemOpen
  \bibfield  {author} {\bibinfo {author} {\bibfnamefont {I.}~\bibnamefont
  {Belopolski}}, \bibinfo {author} {\bibfnamefont {K.}~\bibnamefont {Manna}},
  \bibinfo {author} {\bibfnamefont {D.~S.}\ \bibnamefont {Sanchez}}, \bibinfo
  {author} {\bibfnamefont {G.}~\bibnamefont {Chang}}, \bibinfo {author}
  {\bibfnamefont {B.}~\bibnamefont {Ernst}}, \bibinfo {author} {\bibfnamefont
  {J.}~\bibnamefont {Yin}}, \bibinfo {author} {\bibfnamefont {S.~S.}\
  \bibnamefont {Zhang}}, \bibinfo {author} {\bibfnamefont {T.}~\bibnamefont
  {Cochran}}, \bibinfo {author} {\bibfnamefont {N.}~\bibnamefont {Shumiya}},
  \bibinfo {author} {\bibfnamefont {H.}~\bibnamefont {Zheng}}, \bibinfo
  {author} {\bibfnamefont {B.}~\bibnamefont {Singh}}, \bibinfo {author}
  {\bibfnamefont {G.}~\bibnamefont {Bian}}, \bibinfo {author} {\bibfnamefont
  {D.}~\bibnamefont {Multer}}, \bibinfo {author} {\bibfnamefont
  {M.}~\bibnamefont {Litskevich}}, \bibinfo {author} {\bibfnamefont
  {X.}~\bibnamefont {Zhou}}, \bibinfo {author} {\bibfnamefont {S.-M.}\
  \bibnamefont {Huang}}, \bibinfo {author} {\bibfnamefont {B.}~\bibnamefont
  {Wang}}, \bibinfo {author} {\bibfnamefont {T.-R.}\ \bibnamefont {Chang}},
  \bibinfo {author} {\bibfnamefont {S.-Y.}\ \bibnamefont {Xu}}, \bibinfo
  {author} {\bibfnamefont {A.}~\bibnamefont {Bansil}}, \bibinfo {author}
  {\bibfnamefont {C.}~\bibnamefont {Felser}}, \bibinfo {author} {\bibfnamefont
  {H.}~\bibnamefont {Lin}},\ and\ \bibinfo {author} {\bibfnamefont {M.~Z.}\
  \bibnamefont {Hasan}},\ }\href {https://doi.org/10.1126/science.aav2327}
  {\bibfield  {journal} {\bibinfo  {journal} {Science}\ }\textbf {\bibinfo
  {volume} {365}},\ \bibinfo {pages} {1278} (\bibinfo {year}
  {2019})}\BibitemShut {NoStop}%
\bibitem [{\citenamefont {Liu}\ \emph {et~al.}(2019)\citenamefont {Liu},
  \citenamefont {Liang}, \citenamefont {Liu}, \citenamefont {Xu}, \citenamefont
  {Li}, \citenamefont {Chen}, \citenamefont {Pei}, \citenamefont {Shi},
  \citenamefont {Mo}, \citenamefont {Dudin}, \citenamefont {Kim}, \citenamefont
  {Cacho}, \citenamefont {Li}, \citenamefont {Sun}, \citenamefont {Yang},
  \citenamefont {Liu}, \citenamefont {Parkin}, \citenamefont {Felser},\ and\
  \citenamefont {Chen}}]{Liu1282}%
  \BibitemOpen
  \bibfield  {author} {\bibinfo {author} {\bibfnamefont {D.~F.}\ \bibnamefont
  {Liu}}, \bibinfo {author} {\bibfnamefont {A.~J.}\ \bibnamefont {Liang}},
  \bibinfo {author} {\bibfnamefont {E.~K.}\ \bibnamefont {Liu}}, \bibinfo
  {author} {\bibfnamefont {Q.~N.}\ \bibnamefont {Xu}}, \bibinfo {author}
  {\bibfnamefont {Y.~W.}\ \bibnamefont {Li}}, \bibinfo {author} {\bibfnamefont
  {C.}~\bibnamefont {Chen}}, \bibinfo {author} {\bibfnamefont {D.}~\bibnamefont
  {Pei}}, \bibinfo {author} {\bibfnamefont {W.~J.}\ \bibnamefont {Shi}},
  \bibinfo {author} {\bibfnamefont {S.~K.}\ \bibnamefont {Mo}}, \bibinfo
  {author} {\bibfnamefont {P.}~\bibnamefont {Dudin}}, \bibinfo {author}
  {\bibfnamefont {T.}~\bibnamefont {Kim}}, \bibinfo {author} {\bibfnamefont
  {C.}~\bibnamefont {Cacho}}, \bibinfo {author} {\bibfnamefont
  {G.}~\bibnamefont {Li}}, \bibinfo {author} {\bibfnamefont {Y.}~\bibnamefont
  {Sun}}, \bibinfo {author} {\bibfnamefont {L.~X.}\ \bibnamefont {Yang}},
  \bibinfo {author} {\bibfnamefont {Z.~K.}\ \bibnamefont {Liu}}, \bibinfo
  {author} {\bibfnamefont {S.~S.~P.}\ \bibnamefont {Parkin}}, \bibinfo {author}
  {\bibfnamefont {C.}~\bibnamefont {Felser}},\ and\ \bibinfo {author}
  {\bibfnamefont {Y.~L.}\ \bibnamefont {Chen}},\ }\href
  {https://doi.org/10.1126/science.aav2873} {\bibfield  {journal} {\bibinfo
  {journal} {Science}\ }\textbf {\bibinfo {volume} {365}},\ \bibinfo {pages}
  {1282} (\bibinfo {year} {2019})}\BibitemShut {NoStop}%
\bibitem [{\citenamefont {Liu}\ \emph {et~al.}(2021)\citenamefont {Liu},
  \citenamefont {Fang}, \citenamefont {Fu}, \citenamefont {Ge}, \citenamefont
  {Kareev}, \citenamefont {Kim}, \citenamefont {Choi}, \citenamefont
  {Karapetrova}, \citenamefont {Zhang}, \citenamefont {Gu}, \citenamefont
  {Choi}, \citenamefont {Wen}, \citenamefont {Wilson}, \citenamefont {Fabbris},
  \citenamefont {Ryan}, \citenamefont {Freeland}, \citenamefont {Haskel},
  \citenamefont {Wu}, \citenamefont {Pixley},\ and\ \citenamefont
  {Chakhalian}}]{PhysRevLett.127.277204}%
  \BibitemOpen
  \bibfield  {author} {\bibinfo {author} {\bibfnamefont {X.}~\bibnamefont
  {Liu}}, \bibinfo {author} {\bibfnamefont {S.}~\bibnamefont {Fang}}, \bibinfo
  {author} {\bibfnamefont {Y.}~\bibnamefont {Fu}}, \bibinfo {author}
  {\bibfnamefont {W.}~\bibnamefont {Ge}}, \bibinfo {author} {\bibfnamefont
  {M.}~\bibnamefont {Kareev}}, \bibinfo {author} {\bibfnamefont {J.-W.}\
  \bibnamefont {Kim}}, \bibinfo {author} {\bibfnamefont {Y.}~\bibnamefont
  {Choi}}, \bibinfo {author} {\bibfnamefont {E.}~\bibnamefont {Karapetrova}},
  \bibinfo {author} {\bibfnamefont {Q.}~\bibnamefont {Zhang}}, \bibinfo
  {author} {\bibfnamefont {L.}~\bibnamefont {Gu}}, \bibinfo {author}
  {\bibfnamefont {E.-S.}\ \bibnamefont {Choi}}, \bibinfo {author}
  {\bibfnamefont {F.}~\bibnamefont {Wen}}, \bibinfo {author} {\bibfnamefont
  {J.~H.}\ \bibnamefont {Wilson}}, \bibinfo {author} {\bibfnamefont
  {G.}~\bibnamefont {Fabbris}}, \bibinfo {author} {\bibfnamefont {P.~J.}\
  \bibnamefont {Ryan}}, \bibinfo {author} {\bibfnamefont {J.~W.}\ \bibnamefont
  {Freeland}}, \bibinfo {author} {\bibfnamefont {D.}~\bibnamefont {Haskel}},
  \bibinfo {author} {\bibfnamefont {W.}~\bibnamefont {Wu}}, \bibinfo {author}
  {\bibfnamefont {J.~H.}\ \bibnamefont {Pixley}},\ and\ \bibinfo {author}
  {\bibfnamefont {J.}~\bibnamefont {Chakhalian}},\ }\href
  {https://doi.org/10.1103/PhysRevLett.127.277204} {\bibfield  {journal}
  {\bibinfo  {journal} {Phys. Rev. Lett.}\ }\textbf {\bibinfo {volume} {127}},\
  \bibinfo {pages} {277204} (\bibinfo {year} {2021})}\BibitemShut {NoStop}%
\bibitem [{\citenamefont {Shirley}(1965)}]{Shirley}%
  \BibitemOpen
  \bibfield  {author} {\bibinfo {author} {\bibfnamefont {J.~H.}\ \bibnamefont
  {Shirley}},\ }\href {https://doi.org/10.1103/PhysRev.138.B979} {\bibfield
  {journal} {\bibinfo  {journal} {Phys. Rev.}\ }\textbf {\bibinfo {volume}
  {138}},\ \bibinfo {pages} {B979} (\bibinfo {year} {1965})}\BibitemShut
  {NoStop}%
\bibitem [{\citenamefont {Goldman}\ and\ \citenamefont
  {Dalibard}(2014)}]{Goldman}%
  \BibitemOpen
  \bibfield  {author} {\bibinfo {author} {\bibfnamefont {N.}~\bibnamefont
  {Goldman}}\ and\ \bibinfo {author} {\bibfnamefont {J.}~\bibnamefont
  {Dalibard}},\ }\href {https://doi.org/10.1103/PhysRevX.4.031027} {\bibfield
  {journal} {\bibinfo  {journal} {Phys. Rev. X}\ }\textbf {\bibinfo {volume}
  {4}},\ \bibinfo {pages} {031027} (\bibinfo {year} {2014})}\BibitemShut
  {NoStop}%
\bibitem [{\citenamefont {Eckardt}\ and\ \citenamefont
  {Anisimovas}(2015)}]{Eckardt2015}%
  \BibitemOpen
  \bibfield  {author} {\bibinfo {author} {\bibfnamefont {A.}~\bibnamefont
  {Eckardt}}\ and\ \bibinfo {author} {\bibfnamefont {E.}~\bibnamefont
  {Anisimovas}},\ }\href {https://doi.org/10.1088/1367-2630/17/9/093039}
  {\bibfield  {journal} {\bibinfo  {journal} {New J. Phys.}\ }\textbf {\bibinfo
  {volume} {17}},\ \bibinfo {pages} {093039} (\bibinfo {year}
  {2015})}\BibitemShut {NoStop}%
\bibitem [{\citenamefont {Bukov}\ \emph {et~al.}(2015)\citenamefont {Bukov},
  \citenamefont {D'Alessio},\ and\ \citenamefont {Polkovnikov}}]{Bukov2015}%
  \BibitemOpen
  \bibfield  {author} {\bibinfo {author} {\bibfnamefont {M.}~\bibnamefont
  {Bukov}}, \bibinfo {author} {\bibfnamefont {L.}~\bibnamefont {D'Alessio}},\
  and\ \bibinfo {author} {\bibfnamefont {A.}~\bibnamefont {Polkovnikov}},\
  }\href {https://doi.org/10.1080/00018732.2015.1055918} {\bibfield  {journal}
  {\bibinfo  {journal} {Adv. Phys.}\ }\textbf {\bibinfo {volume} {64}},\
  \bibinfo {pages} {139} (\bibinfo {year} {2015})}\BibitemShut {NoStop}%
\bibitem [{\citenamefont {Oka}\ and\ \citenamefont {Kitamura}(2019)}]{Oka2019}%
  \BibitemOpen
  \bibfield  {author} {\bibinfo {author} {\bibfnamefont {T.}~\bibnamefont
  {Oka}}\ and\ \bibinfo {author} {\bibfnamefont {S.}~\bibnamefont {Kitamura}},\
  }\href {https://doi.org/10.1146/annurev-conmatphys-031218-013423} {\bibfield
  {journal} {\bibinfo  {journal} {Annu. Rev. Condens. Matter Phys.}\ }\textbf
  {\bibinfo {volume} {10}},\ \bibinfo {pages} {387} (\bibinfo {year}
  {2019})}\BibitemShut {NoStop}%
\bibitem [{\citenamefont {Bao}\ \emph {et~al.}(2022)\citenamefont {Bao},
  \citenamefont {Tang}, \citenamefont {Sun},\ and\ \citenamefont
  {Zhou}}]{Light2022}%
  \BibitemOpen
  \bibfield  {author} {\bibinfo {author} {\bibfnamefont {C.}~\bibnamefont
  {Bao}}, \bibinfo {author} {\bibfnamefont {P.}~\bibnamefont {Tang}}, \bibinfo
  {author} {\bibfnamefont {D.}~\bibnamefont {Sun}},\ and\ \bibinfo {author}
  {\bibfnamefont {S.}~\bibnamefont {Zhou}},\ }\href
  {https://doi.org/10.1038/s42254-021-00388-1} {\bibfield  {journal} {\bibinfo
  {journal} {Nat. Rev. Phys.}\ }\textbf {\bibinfo {volume} {4}},\ \bibinfo
  {pages} {33} (\bibinfo {year} {2022})}\BibitemShut {NoStop}%
\bibitem [{\citenamefont {Wang}\ \emph {et~al.}(2013)\citenamefont {Wang},
  \citenamefont {Steinberg}, \citenamefont {Jarillo-Herrero},\ and\
  \citenamefont {Gedik}}]{exp1}%
  \BibitemOpen
  \bibfield  {author} {\bibinfo {author} {\bibfnamefont {Y.}~\bibnamefont
  {Wang}}, \bibinfo {author} {\bibfnamefont {H.}~\bibnamefont {Steinberg}},
  \bibinfo {author} {\bibfnamefont {P.}~\bibnamefont {Jarillo-Herrero}},\ and\
  \bibinfo {author} {\bibfnamefont {N.}~\bibnamefont {Gedik}},\ }\href
  {https://doi.org/10.1126/science.1239834} {\bibfield  {journal} {\bibinfo
  {journal} {Science}\ }\textbf {\bibinfo {volume} {342}},\ \bibinfo {pages}
  {453} (\bibinfo {year} {2013})}\BibitemShut {NoStop}%
\bibitem [{\citenamefont {Mahmood}\ \emph {et~al.}(2016)\citenamefont
  {Mahmood}, \citenamefont {Chan}, \citenamefont {Alpichshev}, \citenamefont
  {Gardner}, \citenamefont {Lee}, \citenamefont {Lee},\ and\ \citenamefont
  {Gedik}}]{exp2}%
  \BibitemOpen
  \bibfield  {author} {\bibinfo {author} {\bibfnamefont {F.}~\bibnamefont
  {Mahmood}}, \bibinfo {author} {\bibfnamefont {C.-K.}\ \bibnamefont {Chan}},
  \bibinfo {author} {\bibfnamefont {Z.}~\bibnamefont {Alpichshev}}, \bibinfo
  {author} {\bibfnamefont {D.}~\bibnamefont {Gardner}}, \bibinfo {author}
  {\bibfnamefont {Y.}~\bibnamefont {Lee}}, \bibinfo {author} {\bibfnamefont
  {P.~A.}\ \bibnamefont {Lee}},\ and\ \bibinfo {author} {\bibfnamefont
  {N.}~\bibnamefont {Gedik}},\ }\href
  {https://doi.org/https://doi.org/10.1038/nphys3609} {\bibfield  {journal}
  {\bibinfo  {journal} {Nat. Phys.}\ }\textbf {\bibinfo {volume} {12}},\
  \bibinfo {pages} {306} (\bibinfo {year} {2016})}\BibitemShut {NoStop}%
\bibitem [{\citenamefont {McIver}\ \emph {et~al.}(2020)\citenamefont {McIver},
  \citenamefont {Schulte}, \citenamefont {Stein}, \citenamefont {Matsuyama},
  \citenamefont {Jotzu}, \citenamefont {Meier},\ and\ \citenamefont
  {Cavalleri}}]{exp3}%
  \BibitemOpen
  \bibfield  {author} {\bibinfo {author} {\bibfnamefont {J.~W.}\ \bibnamefont
  {McIver}}, \bibinfo {author} {\bibfnamefont {B.}~\bibnamefont {Schulte}},
  \bibinfo {author} {\bibfnamefont {F.-U.}\ \bibnamefont {Stein}}, \bibinfo
  {author} {\bibfnamefont {T.}~\bibnamefont {Matsuyama}}, \bibinfo {author}
  {\bibfnamefont {G.}~\bibnamefont {Jotzu}}, \bibinfo {author} {\bibfnamefont
  {G.}~\bibnamefont {Meier}},\ and\ \bibinfo {author} {\bibfnamefont
  {A.}~\bibnamefont {Cavalleri}},\ }\href
  {https://doi.org/https://doi.org/10.1038/s41567-019-0698-y} {\bibfield
  {journal} {\bibinfo  {journal} {Nat. Phys.}\ }\textbf {\bibinfo {volume}
  {16}},\ \bibinfo {pages} {38} (\bibinfo {year} {2020})}\BibitemShut {NoStop}%
\bibitem [{\citenamefont {Oka}\ and\ \citenamefont {Aoki}(2009)}]{FT1}%
  \BibitemOpen
  \bibfield  {author} {\bibinfo {author} {\bibfnamefont {T.}~\bibnamefont
  {Oka}}\ and\ \bibinfo {author} {\bibfnamefont {H.}~\bibnamefont {Aoki}},\
  }\href {https://doi.org/10.1103/PhysRevB.79.081406} {\bibfield  {journal}
  {\bibinfo  {journal} {Phys. Rev. B}\ }\textbf {\bibinfo {volume} {79}},\
  \bibinfo {pages} {081406} (\bibinfo {year} {2009})}\BibitemShut {NoStop}%
\bibitem [{\citenamefont {Lindner}\ \emph {et~al.}(2011)\citenamefont
  {Lindner}, \citenamefont {Refael},\ and\ \citenamefont {Galitski}}]{FT2}%
  \BibitemOpen
  \bibfield  {author} {\bibinfo {author} {\bibfnamefont {N.~H.}\ \bibnamefont
  {Lindner}}, \bibinfo {author} {\bibfnamefont {G.}~\bibnamefont {Refael}},\
  and\ \bibinfo {author} {\bibfnamefont {V.}~\bibnamefont {Galitski}},\ }\href
  {https://doi.org/https://doi.org/10.1038/nphys1926} {\bibfield  {journal}
  {\bibinfo  {journal} {Nat. Phys.}\ }\textbf {\bibinfo {volume} {7}},\
  \bibinfo {pages} {490} (\bibinfo {year} {2011})}\BibitemShut {NoStop}%
\bibitem [{\citenamefont {Harper}\ \emph {et~al.}(2020)\citenamefont {Harper},
  \citenamefont {Roy}, \citenamefont {Rudner},\ and\ \citenamefont
  {Sondhi}}]{Rev1}%
  \BibitemOpen
  \bibfield  {author} {\bibinfo {author} {\bibfnamefont {F.}~\bibnamefont
  {Harper}}, \bibinfo {author} {\bibfnamefont {R.}~\bibnamefont {Roy}},
  \bibinfo {author} {\bibfnamefont {M.~S.}\ \bibnamefont {Rudner}},\ and\
  \bibinfo {author} {\bibfnamefont {S.}~\bibnamefont {Sondhi}},\ }\href
  {https://doi.org/10.1146/annurev-conmatphys-031218-013721} {\bibfield
  {journal} {\bibinfo  {journal} {Annu. Rev. Condens. Matter Phys.}\ }\textbf
  {\bibinfo {volume} {11}},\ \bibinfo {pages} {345} (\bibinfo {year}
  {2020})}\BibitemShut {NoStop}%
\bibitem [{\citenamefont {Rudner}\ and\ \citenamefont {Lindner}(2020)}]{Rev2}%
  \BibitemOpen
  \bibfield  {author} {\bibinfo {author} {\bibfnamefont {M.~S.}\ \bibnamefont
  {Rudner}}\ and\ \bibinfo {author} {\bibfnamefont {N.~H.}\ \bibnamefont
  {Lindner}},\ }\href {https://doi.org/10.1038/s42254-020-0170-z} {\bibfield
  {journal} {\bibinfo  {journal} {Nat. Rev. Phys.}\ }\textbf {\bibinfo {volume}
  {2}},\ \bibinfo {pages} {229} (\bibinfo {year} {2020})}\BibitemShut {NoStop}%
\bibitem [{\citenamefont {Bao}\ \emph {et~al.}(2021)\citenamefont {Bao},
  \citenamefont {Tang}, \citenamefont {Sun},\ and\ \citenamefont
  {Zhou}}]{Rev3}%
  \BibitemOpen
  \bibfield  {author} {\bibinfo {author} {\bibfnamefont {C.}~\bibnamefont
  {Bao}}, \bibinfo {author} {\bibfnamefont {P.}~\bibnamefont {Tang}}, \bibinfo
  {author} {\bibfnamefont {D.}~\bibnamefont {Sun}},\ and\ \bibinfo {author}
  {\bibfnamefont {S.}~\bibnamefont {Zhou}},\ }\href
  {https://doi.org/10.1038/s42254-021-00388-1} {\bibfield  {journal} {\bibinfo
  {journal} {Nat. Rev. Phys.}\ ,\ \bibinfo {pages} {1}} (\bibinfo {year}
  {2021})}\BibitemShut {NoStop}%
\bibitem [{\citenamefont {H{\"u}bener}\ \emph {et~al.}(2017)\citenamefont
  {H{\"u}bener}, \citenamefont {Sentef}, \citenamefont {De~Giovannini},
  \citenamefont {Kemper},\ and\ \citenamefont {Rubio}}]{FWSM1}%
  \BibitemOpen
  \bibfield  {author} {\bibinfo {author} {\bibfnamefont {H.}~\bibnamefont
  {H{\"u}bener}}, \bibinfo {author} {\bibfnamefont {M.~A.}\ \bibnamefont
  {Sentef}}, \bibinfo {author} {\bibfnamefont {U.}~\bibnamefont
  {De~Giovannini}}, \bibinfo {author} {\bibfnamefont {A.~F.}\ \bibnamefont
  {Kemper}},\ and\ \bibinfo {author} {\bibfnamefont {A.}~\bibnamefont
  {Rubio}},\ }\href {https://doi.org/https://doi.org/10.1038/ncomms13940}
  {\bibfield  {journal} {\bibinfo  {journal} {Nat. Commun.}\ }\textbf {\bibinfo
  {volume} {8}},\ \bibinfo {pages} {13940} (\bibinfo {year}
  {2017})}\BibitemShut {NoStop}%
\bibitem [{\citenamefont {Narayan}(2016)}]{FWSM2}%
  \BibitemOpen
  \bibfield  {author} {\bibinfo {author} {\bibfnamefont {A.}~\bibnamefont
  {Narayan}},\ }\href {https://doi.org/10.1103/PhysRevB.94.041409} {\bibfield
  {journal} {\bibinfo  {journal} {Phys. Rev. B}\ }\textbf {\bibinfo {volume}
  {94}},\ \bibinfo {pages} {041409} (\bibinfo {year} {2016})}\BibitemShut
  {NoStop}%
\bibitem [{\citenamefont {Yan}\ and\ \citenamefont {Wang}(2016)}]{FWSM3}%
  \BibitemOpen
  \bibfield  {author} {\bibinfo {author} {\bibfnamefont {Z.}~\bibnamefont
  {Yan}}\ and\ \bibinfo {author} {\bibfnamefont {Z.}~\bibnamefont {Wang}},\
  }\href {https://doi.org/10.1103/PhysRevLett.117.087402} {\bibfield  {journal}
  {\bibinfo  {journal} {Phys. Rev. Lett.}\ }\textbf {\bibinfo {volume} {117}},\
  \bibinfo {pages} {087402} (\bibinfo {year} {2016})}\BibitemShut {NoStop}%
\bibitem [{\citenamefont {Taguchi}\ \emph {et~al.}(2016)\citenamefont
  {Taguchi}, \citenamefont {Xu}, \citenamefont {Yamakage},\ and\ \citenamefont
  {Law}}]{FWSM4}%
  \BibitemOpen
  \bibfield  {author} {\bibinfo {author} {\bibfnamefont {K.}~\bibnamefont
  {Taguchi}}, \bibinfo {author} {\bibfnamefont {D.-H.}\ \bibnamefont {Xu}},
  \bibinfo {author} {\bibfnamefont {A.}~\bibnamefont {Yamakage}},\ and\
  \bibinfo {author} {\bibfnamefont {K.~T.}\ \bibnamefont {Law}},\ }\href
  {https://doi.org/10.1103/PhysRevB.94.155206} {\bibfield  {journal} {\bibinfo
  {journal} {Phys. Rev. B}\ }\textbf {\bibinfo {volume} {94}},\ \bibinfo
  {pages} {155206} (\bibinfo {year} {2016})}\BibitemShut {NoStop}%
\bibitem [{\citenamefont {Gonz\'alez}\ and\ \citenamefont
  {Molina}(2016)}]{FWSM5}%
  \BibitemOpen
  \bibfield  {author} {\bibinfo {author} {\bibfnamefont {J.}~\bibnamefont
  {Gonz\'alez}}\ and\ \bibinfo {author} {\bibfnamefont {R.~A.}\ \bibnamefont
  {Molina}},\ }\href {https://doi.org/10.1103/PhysRevLett.116.156803}
  {\bibfield  {journal} {\bibinfo  {journal} {Phys. Rev. Lett.}\ }\textbf
  {\bibinfo {volume} {116}},\ \bibinfo {pages} {156803} (\bibinfo {year}
  {2016})}\BibitemShut {NoStop}%
\bibitem [{\citenamefont {Chan}\ \emph {et~al.}(2016)\citenamefont {Chan},
  \citenamefont {Oh}, \citenamefont {Han},\ and\ \citenamefont {Lee}}]{FWSM6}%
  \BibitemOpen
  \bibfield  {author} {\bibinfo {author} {\bibfnamefont {C.-K.}\ \bibnamefont
  {Chan}}, \bibinfo {author} {\bibfnamefont {Y.-T.}\ \bibnamefont {Oh}},
  \bibinfo {author} {\bibfnamefont {J.~H.}\ \bibnamefont {Han}},\ and\ \bibinfo
  {author} {\bibfnamefont {P.~A.}\ \bibnamefont {Lee}},\ }\href
  {https://doi.org/10.1103/PhysRevB.94.121106} {\bibfield  {journal} {\bibinfo
  {journal} {Phys. Rev. B}\ }\textbf {\bibinfo {volume} {94}},\ \bibinfo
  {pages} {121106} (\bibinfo {year} {2016})}\BibitemShut {NoStop}%
\bibitem [{\citenamefont {Trevisan}\ \emph {et~al.}(2022)\citenamefont
  {Trevisan}, \citenamefont {Arribi}, \citenamefont {Heinonen}, \citenamefont
  {Slager},\ and\ \citenamefont {Orth}}]{FWSM7}%
  \BibitemOpen
  \bibfield  {author} {\bibinfo {author} {\bibfnamefont {T.~V.}\ \bibnamefont
  {Trevisan}}, \bibinfo {author} {\bibfnamefont {P.~V.}\ \bibnamefont
  {Arribi}}, \bibinfo {author} {\bibfnamefont {O.}~\bibnamefont {Heinonen}},
  \bibinfo {author} {\bibfnamefont {R.-J.}\ \bibnamefont {Slager}},\ and\
  \bibinfo {author} {\bibfnamefont {P.~P.}\ \bibnamefont {Orth}},\ }\href
  {https://doi.org/10.1103/PhysRevLett.128.066602} {\bibfield  {journal}
  {\bibinfo  {journal} {Phys. Rev. Lett.}\ }\textbf {\bibinfo {volume} {128}},\
  \bibinfo {pages} {066602} (\bibinfo {year} {2022})}\BibitemShut {NoStop}%
\bibitem [{\citenamefont {Chen}\ \emph {et~al.}(2018)\citenamefont {Chen},
  \citenamefont {Zhou},\ and\ \citenamefont {Xu}}]{FWSM8}%
  \BibitemOpen
  \bibfield  {author} {\bibinfo {author} {\bibfnamefont {R.}~\bibnamefont
  {Chen}}, \bibinfo {author} {\bibfnamefont {B.}~\bibnamefont {Zhou}},\ and\
  \bibinfo {author} {\bibfnamefont {D.-H.}\ \bibnamefont {Xu}},\ }\href
  {https://doi.org/10.1103/PhysRevB.97.155152} {\bibfield  {journal} {\bibinfo
  {journal} {Phys. Rev. B}\ }\textbf {\bibinfo {volume} {97}},\ \bibinfo
  {pages} {155152} (\bibinfo {year} {2018})}\BibitemShut {NoStop}%
\bibitem [{\citenamefont {Yan}\ and\ \citenamefont {Wang}(2017)}]{FWSM9}%
  \BibitemOpen
  \bibfield  {author} {\bibinfo {author} {\bibfnamefont {Z.}~\bibnamefont
  {Yan}}\ and\ \bibinfo {author} {\bibfnamefont {Z.}~\bibnamefont {Wang}},\
  }\href {https://doi.org/10.1103/PhysRevB.96.041206} {\bibfield  {journal}
  {\bibinfo  {journal} {Phys. Rev. B}\ }\textbf {\bibinfo {volume} {96}},\
  \bibinfo {pages} {041206} (\bibinfo {year} {2017})}\BibitemShut {NoStop}%
\bibitem [{\citenamefont {Zhang}\ \emph {et~al.}(2020)\citenamefont {Zhang},
  \citenamefont {Ding}, \citenamefont {Gan}, \citenamefont {Cao}, \citenamefont
  {Li}, \citenamefont {Wu},\ and\ \citenamefont {Wang}}]{PhysRevB.101.235119}%
  \BibitemOpen
  \bibfield  {author} {\bibinfo {author} {\bibfnamefont {C.}~\bibnamefont
  {Zhang}}, \bibinfo {author} {\bibfnamefont {X.-Y.}\ \bibnamefont {Ding}},
  \bibinfo {author} {\bibfnamefont {L.-Y.}\ \bibnamefont {Gan}}, \bibinfo
  {author} {\bibfnamefont {Y.}~\bibnamefont {Cao}}, \bibinfo {author}
  {\bibfnamefont {B.-S.}\ \bibnamefont {Li}}, \bibinfo {author} {\bibfnamefont
  {X.}~\bibnamefont {Wu}},\ and\ \bibinfo {author} {\bibfnamefont
  {R.}~\bibnamefont {Wang}},\ }\href
  {https://doi.org/10.1103/PhysRevB.101.235119} {\bibfield  {journal} {\bibinfo
   {journal} {Phys. Rev. B}\ }\textbf {\bibinfo {volume} {101}},\ \bibinfo
  {pages} {235119} (\bibinfo {year} {2020})}\BibitemShut {NoStop}%
\bibitem [{\citenamefont {Deng}\ \emph {et~al.}(2020)\citenamefont {Deng},
  \citenamefont {Zheng}, \citenamefont {Zhan}, \citenamefont {Fan},
  \citenamefont {Wu},\ and\ \citenamefont {Wang}}]{PhysRevB.102.201105}%
  \BibitemOpen
  \bibfield  {author} {\bibinfo {author} {\bibfnamefont {T.}~\bibnamefont
  {Deng}}, \bibinfo {author} {\bibfnamefont {B.}~\bibnamefont {Zheng}},
  \bibinfo {author} {\bibfnamefont {F.}~\bibnamefont {Zhan}}, \bibinfo {author}
  {\bibfnamefont {J.}~\bibnamefont {Fan}}, \bibinfo {author} {\bibfnamefont
  {X.}~\bibnamefont {Wu}},\ and\ \bibinfo {author} {\bibfnamefont
  {R.}~\bibnamefont {Wang}},\ }\href
  {https://doi.org/10.1103/PhysRevB.102.201105} {\bibfield  {journal} {\bibinfo
   {journal} {Phys. Rev. B}\ }\textbf {\bibinfo {volume} {102}},\ \bibinfo
  {pages} {201105} (\bibinfo {year} {2020})}\BibitemShut {NoStop}%
\bibitem [{\citenamefont {Hohenberg}\ and\ \citenamefont
  {Kohn}(1964)}]{PhysRev.136.B864}%
  \BibitemOpen
  \bibfield  {author} {\bibinfo {author} {\bibfnamefont {P.}~\bibnamefont
  {Hohenberg}}\ and\ \bibinfo {author} {\bibfnamefont {W.}~\bibnamefont
  {Kohn}},\ }\href {https://doi.org/10.1103/PhysRev.136.B864} {\bibfield
  {journal} {\bibinfo  {journal} {Phys. Rev.}\ }\textbf {\bibinfo {volume}
  {136}},\ \bibinfo {pages} {B864} (\bibinfo {year} {1964})}\BibitemShut
  {NoStop}%
\bibitem [{\citenamefont {Kohn}\ and\ \citenamefont
  {Sham}(1965)}]{PhysRev.140.A1133}%
  \BibitemOpen
  \bibfield  {author} {\bibinfo {author} {\bibfnamefont {W.}~\bibnamefont
  {Kohn}}\ and\ \bibinfo {author} {\bibfnamefont {L.~J.}\ \bibnamefont
  {Sham}},\ }\href {https://doi.org/10.1103/PhysRev.140.A1133} {\bibfield
  {journal} {\bibinfo  {journal} {Phys. Rev.}\ }\textbf {\bibinfo {volume}
  {140}},\ \bibinfo {pages} {A1133} (\bibinfo {year} {1965})}\BibitemShut
  {NoStop}%
\bibitem [{\citenamefont {Kresse}\ and\ \citenamefont
  {Furthm\"uller}(1996)}]{PhysRevB.54.11169}%
  \BibitemOpen
  \bibfield  {author} {\bibinfo {author} {\bibfnamefont {G.}~\bibnamefont
  {Kresse}}\ and\ \bibinfo {author} {\bibfnamefont {J.}~\bibnamefont
  {Furthm\"uller}},\ }\href {https://doi.org/10.1103/PhysRevB.54.11169}
  {\bibfield  {journal} {\bibinfo  {journal} {Phys. Rev. B}\ }\textbf {\bibinfo
  {volume} {54}},\ \bibinfo {pages} {11169} (\bibinfo {year}
  {1996})}\BibitemShut {NoStop}%
\bibitem [{\citenamefont {Perdew}\ \emph {et~al.}(1996)\citenamefont {Perdew},
  \citenamefont {Burke},\ and\ \citenamefont
  {Ernzerhof}}]{PhysRevLett.77.3865}%
  \BibitemOpen
  \bibfield  {author} {\bibinfo {author} {\bibfnamefont {J.~P.}\ \bibnamefont
  {Perdew}}, \bibinfo {author} {\bibfnamefont {K.}~\bibnamefont {Burke}},\ and\
  \bibinfo {author} {\bibfnamefont {M.}~\bibnamefont {Ernzerhof}},\ }\href
  {https://doi.org/10.1103/PhysRevLett.77.3865} {\bibfield  {journal} {\bibinfo
   {journal} {Phys. Rev. Lett.}\ }\textbf {\bibinfo {volume} {77}},\ \bibinfo
  {pages} {3865} (\bibinfo {year} {1996})}\BibitemShut {NoStop}%
\bibitem [{\citenamefont {Marzari}\ \emph {et~al.}(2012)\citenamefont
  {Marzari}, \citenamefont {Mostofi}, \citenamefont {Yates}, \citenamefont
  {Souza},\ and\ \citenamefont {Vanderbilt}}]{Marzari2012}%
  \BibitemOpen
  \bibfield  {author} {\bibinfo {author} {\bibfnamefont {N.}~\bibnamefont
  {Marzari}}, \bibinfo {author} {\bibfnamefont {A.~A.}\ \bibnamefont
  {Mostofi}}, \bibinfo {author} {\bibfnamefont {J.~R.}\ \bibnamefont {Yates}},
  \bibinfo {author} {\bibfnamefont {I.}~\bibnamefont {Souza}},\ and\ \bibinfo
  {author} {\bibfnamefont {D.}~\bibnamefont {Vanderbilt}},\ }\href
  {https://doi.org/10.1103/RevModPhys.84.1419} {\bibfield  {journal} {\bibinfo
  {journal} {Rev. Mod. Phys.}\ }\textbf {\bibinfo {volume} {84}},\ \bibinfo
  {pages} {1419} (\bibinfo {year} {2012})}\BibitemShut {NoStop}%
\bibitem [{\citenamefont {Mostofi}\ \emph {et~al.}(2008)\citenamefont
  {Mostofi}, \citenamefont {Yates}, \citenamefont {Lee}, \citenamefont {Souza},
  \citenamefont {Vanderbilt},\ and\ \citenamefont {Marzari}}]{MOSTOFI2008685}%
  \BibitemOpen
  \bibfield  {author} {\bibinfo {author} {\bibfnamefont {A.~A.}\ \bibnamefont
  {Mostofi}}, \bibinfo {author} {\bibfnamefont {J.~R.}\ \bibnamefont {Yates}},
  \bibinfo {author} {\bibfnamefont {Y.-S.}\ \bibnamefont {Lee}}, \bibinfo
  {author} {\bibfnamefont {I.}~\bibnamefont {Souza}}, \bibinfo {author}
  {\bibfnamefont {D.}~\bibnamefont {Vanderbilt}},\ and\ \bibinfo {author}
  {\bibfnamefont {N.}~\bibnamefont {Marzari}},\ }\href
  {https://doi.org/https://doi.org/10.1016/j.cpc.2007.11.016} {\bibfield
  {journal} {\bibinfo  {journal} {Comput. Phys. Commun.}\ }\textbf {\bibinfo
  {volume} {178}},\ \bibinfo {pages} {685} (\bibinfo {year}
  {2008})}\BibitemShut {NoStop}%
\bibitem [{\citenamefont {Wu}\ \emph {et~al.}(2018)\citenamefont {Wu},
  \citenamefont {Zhang}, \citenamefont {Song}, \citenamefont {Troyer},\ and\
  \citenamefont {Soluyanov}}]{WU2018405}%
  \BibitemOpen
  \bibfield  {author} {\bibinfo {author} {\bibfnamefont {Q.}~\bibnamefont
  {Wu}}, \bibinfo {author} {\bibfnamefont {S.}~\bibnamefont {Zhang}}, \bibinfo
  {author} {\bibfnamefont {H.-F.}\ \bibnamefont {Song}}, \bibinfo {author}
  {\bibfnamefont {M.}~\bibnamefont {Troyer}},\ and\ \bibinfo {author}
  {\bibfnamefont {A.~A.}\ \bibnamefont {Soluyanov}},\ }\href
  {https://doi.org/https://doi.org/10.1016/j.cpc.2017.09.033} {\bibfield
  {journal} {\bibinfo  {journal} {Comput. Phys. Commun.}\ }\textbf {\bibinfo
  {volume} {224}},\ \bibinfo {pages} {405} (\bibinfo {year}
  {2018})}\BibitemShut {NoStop}%
\bibitem [{\citenamefont {Sancho}\ \emph {et~al.}(1984)\citenamefont {Sancho},
  \citenamefont {Sancho},\ and\ \citenamefont {Rubio}}]{Sancho1984}%
  \BibitemOpen
  \bibfield  {author} {\bibinfo {author} {\bibfnamefont {M.~P.~L.}\
  \bibnamefont {Sancho}}, \bibinfo {author} {\bibfnamefont {J.~M.~L.}\
  \bibnamefont {Sancho}},\ and\ \bibinfo {author} {\bibfnamefont
  {J.}~\bibnamefont {Rubio}},\ }\href
  {https://doi.org/10.1088/0305-4608/14/5/016} {\bibfield  {journal} {\bibinfo
  {journal} {J. Phys. F: Metal Phys.}\ }\textbf {\bibinfo {volume} {14}},\
  \bibinfo {pages} {1205} (\bibinfo {year} {1984})}\BibitemShut {NoStop}%
\bibitem [{\citenamefont {Sancho}\ \emph {et~al.}(1985)\citenamefont {Sancho},
  \citenamefont {Sancho}, \citenamefont {Sancho},\ and\ \citenamefont
  {Rubio}}]{RGFM}%
  \BibitemOpen
  \bibfield  {author} {\bibinfo {author} {\bibfnamefont {M.~L.}\ \bibnamefont
  {Sancho}}, \bibinfo {author} {\bibfnamefont {J.~L.}\ \bibnamefont {Sancho}},
  \bibinfo {author} {\bibfnamefont {J.~L.}\ \bibnamefont {Sancho}},\ and\
  \bibinfo {author} {\bibfnamefont {J.}~\bibnamefont {Rubio}},\ }\href
  {https://doi.org/10.1088/0305-4608/15/4/009} {\bibfield  {journal} {\bibinfo
  {journal} {J. Phys. F: Met. Phys.}\ }\textbf {\bibinfo {volume} {15}},\
  \bibinfo {pages} {851} (\bibinfo {year} {1985})}\BibitemShut {NoStop}%
\end{thebibliography}

%apsrev4-2.bst 2019-01-14 (MD) hand-edited version of apsrev4-1.bst
%Control: key (0)
%Control: author (72) initials jnrlst
%Control: editor formatted (1) identically to author
%Control: production of article title (-1) disabled
%Control: page (0) single
%Control: year (1) truncated
%Control: production of eprint (0) enabled
%

\end{document}